\documentclass[10pt,twocolumn,twoside]{IEEEtran}

\usepackage{cite}
\usepackage{graphicx,subfigure}
\usepackage{psfrag}
\usepackage{amsmath,amssymb}
\def\reals{{\mathbb{R}}}
\def\complex{{\mathbb{C}}}
\def\naturals{{\mathbb{N}}}

\def\ints{{\mathbb{Z}}}

\def\bfZero{\boldsymbol{0}}

\def\bfa{{\boldsymbol{a}}}

\def\bfb{{\boldsymbol{b}}}

\def\bfd{{\boldsymbol{d}}}

\def\bfr{{\boldsymbol{r}}}
\def\bfrh{{\hat{\boldsymbol{r}}}}

\def\bfs{{\boldsymbol{s}}}

\def\bfu{{\boldsymbol{u}}}

\def\bfv{{\boldsymbol{v}}}

\def\bfw{{\boldsymbol{w}}}

\def\bfx{{\boldsymbol{x}}}
\def\bfxh{{\hat{\boldsymbol{x}}}}

\def\bfy{{\boldsymbol{y}}}
\def\bfyh{{\hat{\boldsymbol{y}}}}

\def\bfz{{\boldsymbol{z}}}

\def\bfB{{\boldsymbol{B}}}

\def\bfE{{\boldsymbol{E}}}

\def\bfF{{\boldsymbol{F}}}

\def\bfG{{\boldsymbol{G}}}

\def\bfH{{\boldsymbol{H}}}

\def\bfHt{{\tilde{\boldsymbol{H}}}}

\def\bfI{{\boldsymbol{I}}}

\def\bfM{{\boldsymbol{M}}}

\def\bfR{{\boldsymbol{R}}}

\def\bfT{{\boldsymbol{T}}}

\def\bfU{{\boldsymbol{U}}}

\def\bfV{{\boldsymbol{V}}}

\def\bfW{{\boldsymbol{W}}}

\def\bfX{{\boldsymbol{X}}}

\def\bfY{{\boldsymbol{Y}}}

\def\bfSigma{{\boldsymbol{\Sigma}}}

\def\Aset{\mathcal{A}}

\def\Bset{\mathcal{B}}

\def\Dset{\mathcal{D}}

\def\Oset{\mathcal{O}}

\def\Rset{\mathcal{R}}

\def\Tset{\mathcal{T}}

\def\Vset{\mathcal{V}}

\def\Xset{\mathcal{X}}

\def\tr{^{\mathrm T}}
\def\hr{^{\mathrm H}}

\def\fro{{\mathrm F}}

\newcommand{\prob}[1]{\mathrm{P}\left(#1\right)}
\newcommand{\expt}[2][]{\mathrm{E}_{#1}\left\{ #2 \right\}}

\DeclareMathOperator*{\dotleq}{\overset{.}{\leq}}
\DeclareMathOperator*{\dotgeq}{\overset{.}{\geq}}
\DeclareMathOperator*{\defeq}{\triangleq}

\newtheorem{theorem}{Theorem}
\newtheorem{corollary}{Corollary}[theorem]
\newtheorem{lemma}{Lemma}
\newtheorem{definition}{Definition}

\def\ml{\mathrm{ML}}
\def\lat{\mathrm{L}}
\def\app{\mathrm{A}}
\def\nlat{\mathrm{NL}}
\def\out{\mathrm{out}}
\def\nt{n_\mathrm{T}}
\def\nr{n_\mathrm{R}}

\def\ints{\mathbb{Z}}

\def\normal{\mathcal{N}}

\def\kron{\otimes}
\def\fro{_\mathrm{F}}
\def\kron{\otimes}
\def\vol{V}

\begin{document}
\sloppy
\title{DMT Optimality of LR-Aided Linear Decoders\\ for a General Class of Channels, Lattice Designs, and System Models}

\author{\thanks{This work was supported by the European Commission in the framework of the FP7 Network of Excellence in Wireless COMmunications NEWCOM++ (contract n. 216715). J. Jald{\'e}n acknowledges funding by FWF Grant N10606 (SISE). The material in this paper will be presented in part at the IEEE International Symposium on Information Theory (ISIT 2009), Seoul, Korea.}
Joakim Jald{\'e}n, \IEEEmembership{Member, IEEE}, \thanks{J. Jald{\'e}n is with the Institute of Communications and Radio-Frequency Engineering, Technical University of Vienna, A-1040 Vienna, Austria (email: joakim.jalden@nt.tuwien.ac.at)}
and Petros Elia, \IEEEmembership{Member, IEEE} \thanks{P. Elia is with the Mobile Communications Department, EURECOM, F-06904 Sophia Antipolis cedex, France (email: elia@eurecom.fr)}
}

\markboth{Submitted to the IEEE Transactions on Information Theory, May 25, 2009}{Submitted to the IEEE Transactions on Information Theory, May 25, 2009}
\maketitle

\begin{abstract}
The work identifies the first general, explicit, and non-random MIMO encoder-decoder structures that guarantee optimality with respect to the diversity-multiplexing tradeoff (DMT), without employing a computationally expensive maximum-likelihood (ML) receiver. Specifically, the work establishes the DMT optimality of a class of regularized lattice decoders, and more importantly the DMT optimality of their lattice-reduction (LR)-aided \emph{linear} counterparts. The results hold for all channel statistics, for all channel dimensions, and most interestingly, irrespective of the particular lattice-code applied. As a special case, it is established that the LLL-based LR-aided linear implementation of the MMSE-GDFE lattice decoder facilitates DMT optimal decoding of any lattice code at a worst-case complexity that grows at most linearly in the data rate. This represents a fundamental reduction in the decoding complexity when compared to ML decoding whose complexity is generally exponential in rate.

The results' generality lends them applicable to a plethora of pertinent communication scenarios such as quasi-static MIMO, MIMO-OFDM, ISI, cooperative-relaying, and MIMO-ARQ channels, in all of which the DMT optimality of the LR-aided linear decoder is guaranteed. The adopted approach yields insight, and motivates further study, into joint transceiver designs with an improved SNR gap to ML decoding.
\end{abstract}

\begin{IEEEkeywords}
Diversity-multiplexing tradeoff, lattice decoding, linear decoding, lattice reduction, regularization, multiple-input multiple-output (MIMO), space-time coders-decoders.
\end{IEEEkeywords}

\section{Introduction}

The general multi-dimensional linear channel model
$$
\bfy=\bfH\bfx+\bfw
$$
adequately represents a plethora of communication system models which utilize multi-dimensional transmit-receive signals for attaining increased rates and reliability in the presence of fading. Such system models include quasi-static MIMO, MIMO-OFDM, ISI, amplify-and-forward (AF), decode-and-forward (DF), and MIMO automatic repeat request (ARQ) models.  Each of the above models introduces its own structure on $\bfH$ and $\bfx$, its own error performance limits, and its own requirements on coding and decoding schemes. Finding general-purpose transceiver structures with (provably) good performance in these scenarios, and with a reasonable computational complexity, is challenging.

\subsection{Background and previous work} \label{sec:background}

Substantial amounts of work have focused on identifying performance criteria and constructing different coding schemes specifically suited to the different system models. For example in the case of the $\nt \times \nr$ quasi-static MIMO channel, we have seen the orthogonal space-time (ST) designs \cite{Ala:98,TJC:99} providing full diversity but doing so only at rates much less than those theoretically possible, codes like V-BLAST \cite{WFG:98} providing full rate MIMO benefits but with much reduced diversity, and codes from the general linear dispersion designs \cite{HH:02} providing full rate benefits but no diversity guarantees for increasing spectral efficiencies.

In outage limited communications systems, the fundamental limits with respect to the spectral efficiency and decoding error probability in the high signal-to-noise ratio (SNR) limit were succinctly characterized by Zheng and Tse's \emph{diversity multiplexing tradeoff} (DMT) \cite{ZT:03}. The tradeoff incorporated several previous performance measures and has been extensively adopted ever since as a benchmark for transceiver design and analysis. The work in \cite{ZT:03} also introduced the notion of DMT optimal designs, i.e., designs capable of achieving the fundamental DMT of the underlying channel (c.f., \cite{ZT:03} or Section~\ref{sec:dmt}).

\subsubsection{Coding} \label{sec:coding}

Towards finding DMT optimal codes, the work in \cite{ZT:03} proved the existence of such codes for the case of the i.i.d.\ Rayleigh fading quasi-static MIMO channel by using ensembles of random Gaussian codes over a finite coding duration, and thus reduced system model dimensionality. Although providing codes of finite length, such a construction is highly impractical given the lack of structure that would allow for practical codeword enumeration and decoding. This issue was addressed in \cite{GCD:04} which, for the same setting, proved the existence of random ensembles of DMT optimal codes that accept a lattice structure. The same work successfully identified the suitability of the lattice framework for MIMO coding problems, and its effect on issues such as that of finding efficient shaping regions for the transmitted signals. However, random lattice designs inherently rely on different lattices for each rate and SNR and, furthermore, do not provide deterministic means by which to identify the lattice generator matrices.

These two issues were conclusively solved in \cite{EKP:05,EKP:06} which first provided practical construction criteria for DMT optimal codes for the quasi-static Rayleigh fading MIMO channel, and then explicitly constructed the first unified family of DMT optimal codes for all channel dimensions. These cyclic division algebra (CDA)-based codes, which were built based on the work of \cite{SRS:03,BR:03,KR:05}, managed to employ for any given number of transmit antennas $\nt$ \emph{a single} lattice generator matrix which is easy to identify.  Furthermore, these codes guarantee DMT optimality for all fading statistics, due to the fact that they satisfy the \emph{approximate universality }criterion of \cite{TV:06}.  Other CDA codes \cite{ORB:06}, and later constructed variants of CDA-based codes \cite{ESK:07,KC:09,HLR:06}, currently perform best among all existing ST codes.  Specifically, the \emph{perfect ST codes} proposed in \cite{ORB:06}, and later extended in \cite{ESK:07}, allow for approximate universality as well as \emph{information losslessness} (c.f., \cite{DTB:02}) for rotationally invariant ST channels. Later work in \cite{KC:09} employed the perfect ST code architecture, together with the lattice space-time (LAST) code framework in \cite{GCD:04}, to provide for an improved shaping region and better performance at lower values of SNR.  Furthermore the work in \cite{HLR:06} drew ST codes from subsets of CDAs that constitute \emph{maximal orders}, which interestingly ensure a better fundamental volume of the corresponding lattice, and better energy efficiency \cite{HLR:06}. The above DMT optimal codes form the basis for modified schemes that DMT optimally apply to different system models \cite{YB:07a,Lu:08,EVA:09,PKE:09,EK:09}.

The codes discussed above have to date only been shown to provide DMT optimality in the presence of an ML decoder\footnote{A notable exception are the random LAST codes in \cite{GCD:04}, as discussed in Section~\ref{sec:redcomp} and throughout the present work.}, and hence decoding complexity has remained the fundamental limitation in obtaining (provably) good decoding error probability performance in a computationally efficient manner. This limitation, roughly speaking, originates from the fact that such codes must in general be drawn, due to enumerability and rate requirements, from lattices whose dimension ``matches'' the inherently high dimension of $\bfH$. On top of that, in all but rare cases, the diversity requirements force code-channel lattices that cannot be decomposed into substantially ``smaller'' and simpler component lattices, without severely sacrificing rate gains.  The high dimensionality, in conjunction with the high spectral efficiency that is envisioned in future telecommunications, introduce prohibitive ML decoding complexity.

\subsubsection{Decoding} \label{sec:decoding}

While sphere decoding (SD) methods \cite{AEV:02,DGC:03,MGD:06}, that perform a limited branch-and-bound type search within a hyper-sphere around the received vector, have been developed to provide ML decoding at reduced average complexity, they remain impractical for dense constellations, low-SNR and ill-conditioned or singular channel realizations \cite{AEV:02,DGC:03,MGD:06,JO:04,JO:05}. This is mainly because they implement an exact solution to a closest vector problem (CVP) for each transmitted codeword.

Substantial interest has been drawn by linear receivers based on the zero-forcing (ZF) or the minimum mean square error (MMSE) criteria, as these receivers avoid exact CVP solutions, and thus allow for simple implementation (c.f., \cite{TV:05} and references therein). An inherent limitation of ZF-based linear receivers is that ill-conditioned channel matrices lead to substantial noise amplification.  This motivated the introduction of MMSE-based linear receivers which can be seen as ZF receivers that take into consideration the presence of additive noise and hence utilize a better-conditioned equivalent channel matrix.  It is the case though that for ill-conditioned channel matrices, both these linear receivers, as well as receivers based on successive interference cancellation (SIC), are for the most part substantially suboptimal, as recent DMT analysis in \cite{KCM:07} reveals.

Notable steps towards better performing efficient receivers included the introduction of lattice-reduction (LR) techniques in \cite{YW:02,WF:03}. Motivated by the fact that ZF is optimal in the presence of orthogonal channels, the work in \cite{YW:02,WF:03} proposed the use of LR methods for better, nearly orthogonal conditioning of the equivalent channel matrix, prior to simple ZF or SIC decoding. This approach was partly validated by simulations (c.f., \cite{MGD:06}) and by analysis as in \cite{TMK:07} which showed that LR-aided ZF decoding can achieve maximal receive diversity for fixed-rate uncoded V-BLAST. LR-aided ZF decoding or naive lattice decoding is, however, not DMT optimal in general \cite{GCD:04,TK:07}. The work in \cite{DGC:03,DGC:04,MGD:06} proposed lattice decoding with MMSE-GDFE pre-processing which is well suited for the case of under-determined or singular channels. Contemporary work on LR-aided decoding in an MMSE pre-processed basis appeared in \cite{WBK:04}. Simulation results indicated that such methods are capable of near-ML performance at a computational complexity that remains low \cite{DGC:03,MGD:06, WBK:04,KC:09}.

\subsubsection{Codes with reduced decoding complexity} \label{sec:redcomp}

Several works focused on providing codes with reduced ML decoding complexity. Such work includes the multi-group decodable codes based on Clifford algebras in \cite{KR:09}, and the codes in \cite{HR:07} for asymmetric ($\nr<\nt)$ quasi-static MIMO channels. Similarly motivated work in \cite{BHV:09} identified existing $2\times 2$ full-rate full-diversity codes for the $2\times 2$ MIMO channel \cite{TK:02,PGG:07,SF:07}, as \emph{fast decodable codes} since they incur reduced sphere decoding complexity by essentially reducing the dimensionality of the search space from $8$ real dimensions to $6$ real dimensions.  This reduction is achieved by linearly combining two Alamouti style \emph{twisted} codes, such that the corresponding QR decomposition employed in SD, yields a sparse $\bfR$ matrix.  The sparseness property was shown to be unique to the case of $\nt=T=2$ where $\nt$ and $T$ denotes the number of transmit antennas and the coding duration respectively, and further extensions to the $4\times 2$ MIMO channel came at the expense of reduced diversity \cite{BHV:09}.

Towards bridging the gap between ML and linear decoders, a hybrid transceiver was proposed in \cite{MS:06} to jointly employ an ML and an \emph{unbiased MMSE-SIC} receiver, on an infinitely long ($T\rightarrow \infty$) D-BLAST style $\nt \times T$ space-time spreading (STS) code with an underlying QAM constellation. This hybrid transceiver allows for \emph{partial} reduction in decoding complexity, and provides DMT optimality with $2 \nt$-dimensional ML decoding (in every time slot).  For the case where $\nr \geq \nt$, a pure ML receiver would generally incur a dimensionality of $2 \nt T$ real symbols.

One of the most fundamentally important steps towards establishing that DMT optimality can be achieved with computationally efficient encoders and decoders was, however, given in \cite{GCD:04}. In the setting of the i.i.d.\ Rayleigh fading quasi-static MIMO channel, it was shown that the random codes from the ensemble proposed in \cite{GCD:04} may be DMT optimally decoded by a lattice decoder (whereby the constellation boundaries are ignored in the decoding process). This was accomplished by the inclusion of the MMSE-GDFE pre-processing step and a random lattice translate. It should, however, be noted that an exact implementation of the MMSE-GDFE lattice decoder still requires the solution to a CVP, which is NP-hard in general \cite{Mic:01}. Currently, except for the Alamouti transceiver structure \cite{Ala:98} over the $2 \times 1$ quasi-static MISO channel, all known DMT optimal explicit, non-random, transceivers employ ML detection, and incur worst-case complexity that is exponential in the data rate.

\subsection{Principal results and outline} \label{sec:results}

The contribution of this work lies in the identification of a large class of scenarios where efficient variants of LR-aided linear lattice decoding, which is a generally suboptimal but computationally advantageous decoding strategy, achieve the diversity of the ML decoder.   The work also presents the first explicit characterization of efficient non-ML encoder-decoder structures that meet the fundamental DMT performance limits, for very general channel statistics, dimensions, and models.  DMT optimality is shown to be achieved with the smallest known complexity order among all DMT-optimal decoders that apply to general lattice designs.

As a first step towards providing computationally efficient DMT optimality, Theorem~\ref{thrm:main} in Section~\ref{sec:main}, proves that \emph{regularized lattice decoders} are DMT optimal. The proposed class of decoders employs an \emph{unconstrained lattice search} in a \emph{regularized} metric which applies an incremental penalization to lattice points further from the origin.  The decoder structure includes, as a special case, the MMSE-GDFE lattice decoder \cite{DGC:03,GCD:04}.  The DMT optimality holds irrespective of the channel's fading statistics and irrespective of the lattice design which is decoded (c.f., \cite{EKP:05,EKP:06,SRS:03,BR:03,KR:05,ORB:06,ESK:07,KC:09,HLR:06,YB:07a,Lu:08,EVA:09,PKE:09,EK:09}), as long as the lattice design and fading distribution jointly induce a (right) continuous\footnote{A similar continuity assumption is required (although not explicitly stated) in establishing the DMT optimality of approximately universal codes, c.f., \cite[Th. 3.1]{TV:06}.} DMT curve (c.f., \cite{ZT:03}) under ML detection. Currently all known DMT curves for the system models considered herein are continuous except possibly at the maximal multiplexing gain. The result holds also when ML decoding, due to suboptimality of the code applied, does not achieve the fundamental DMT of the channel. This further strengthens the view of regularized lattice decoding as a DMT optimal \emph{decoding} strategy.

As a second step towards computationally efficient DMT optimality, Theorem~\ref{thrm:approx}, in Section~\ref{sec:c-approx}, extends the above result to the class of all \emph{$C$-approximate} implementations of regularized lattice decoders. Two decoders are here said to be $C$-approximate when their minimum metrics are at a distance less than some \emph{constant} $C$ (c.f., Section~\ref{sec:c-approx}). The DMT optimality of LLL-based LR-aided linear decoders, being $C$-approximate decoders, is then established by Corollary~\ref{cor:lr-aided}.

Theorem~\ref{thrm:complex}, in Section~\ref{sec:lll}, then considers the computational complexity of the LR-aided solutions and proves that LR-aided DMT optimal decoding is feasible at a \emph{worst-case} complexity of $O(\log \rho)$ where $\rho$ denotes the SNR, i.e., at a complexity which grows only linearly in the data rate. With LLL LR worst-case complexity known to be generally unbounded \cite{JSM:08}, the upper bound is guaranteed by exploiting channel information at the receiver and rigorously relating lattices that result in high probability of error, to lattices that may induce high LR complexity.
The bound quantifies, in the scale of interest, the fundamental reduction in the decoding complexity of the proposed explicit transceivers, when compared to the ML decoder which has a complexity that is generally exponential in the rate.
It also resolves, in the negative, the long standing open problem of whether DMT optimality requires a complexity that is exponential in rate.

Section~\ref{sec:general} considers different generalizations including the case of nested lattice designs, partial channel knowledge, general and possibly non-Gaussian noise characteristics, and provides a discussion of the case where the diversity multiplexing characteristic of some scenario is discontinuous and/or unknown. Section~\ref{sec:examples} then shows how the result directly applies to several pertinent computationally demanding communication scenarios such as MIMO-OFDM, ISI, amplify-and-forward, decode-and-forward and MIMO-ARQ settings, in all of which the DMT optimality of the efficient decoders is guaranteed, again for any lattice design and fading distribution. Conclusions are provided in Section~\ref{sec:conclusion}.

\subsection{Notation}

$\ints,\reals$ and $\complex$ respectively denote the integer, the real and the complex numbers. $\reals^n$ and $\reals^{m \times n}$ denote the set of $n$-dimensional and $m \times n$-dimensional real vectors and matrices.  Similar definitions apply to $\ints$ and $\complex$. Vectors and matrices are respectively denoted by lower- and upper-case bold letters, i.e., $\bfx$ and $\bfX$. The identity matrix is denoted $\bfI$ and its size is made clear by the context. The all-zeros vector or matrix is denoted $\bfZero$. $\bfX\tr$, $\bfX\hr$ and $\bfX^{-1}$ denotes the transpose, conjugate transpose and inverse of a matrix $\bfX$. $\| \bfx \|$ denotes the Euclidean norm of $\bfx$, and $\| \bfX \|\fro^2$ the Frobenius norm of $\bfX$. No notational difference is made between random variables (vectors and matrices) and their realizations. The multivariate real valued Gaussian distribution with zero mean and covariance $\bfI$ is denoted $\normal(\bfZero,\bfI)$.

\section{System model}
\label{sec:model}

\subsection{The generic MIMO channel} \label{sec:channels}

We consider a generic $n \times m$ (real) MIMO channel model
\begin{equation} \label{eq:model}
\bfy = \bfH \bfx + \bfw
\end{equation}
where $\bfy \in \reals^m$, $\bfH \in \reals^{m \times n}$, $\bfx \in \reals^n$ and $\bfw \in \reals^m$. The transmitted codewords $\bfx$ are assumed to be uniformly distributed over some codebook $\Xset \subset \reals^n$, and statistically independent of $\bfH$. The noise is assumed to be i.i.d.\ Gaussian with unit variance, i.e., $\bfw \sim \normal(\bfZero, \bfI)$. Under these assumptions the optimal decoder, in the sense that it minimizes the probability of codeword error, is the ML decoder given by
\begin{equation} \label{eq:ml}
\bfxh_{\ml} = \arg \min_{\bfxh \in \Xset} \|Ê\bfy - \bfH \bfxh \|^2 \, .
\end{equation}

The channel $\bfH$ is assumed random (i.e., fading) with a distribution parameterized by a real parameter $\rho \geq 0$. The parameter $\rho$ will throughout be interpreted as the SNR of the channel, although this is strictly speaking not required for the analysis. We assume that one use of \eqref{eq:model} corresponds to $T$ uses of some underlying ``physical'' channel, which motivates a definition of the rate in terms of bits per channel use (bpcu) according to
\begin{equation} \label{eq:rate}
R = \frac{1}{T} \log_{2} |\Xset|
\end{equation}
where $|\Xset|$ denotes the cardinality or size of $\Xset$. The model in \eqref{eq:model} is known to encompass many pertinent communication scenarios (c.f., \cite{DGC:03}), and several explicit examples are provided in Section~\ref{sec:examples}. The obtained results hold in the general setting unless otherwise explicitly stated.

\subsection{The diversity-multiplexing tradeoff} \label{sec:dmt}

Following \cite{ZT:03} we refer to a family of codes, $\Xset(\rho)$,  parameterized by $\rho$ as a \emph{scheme} and define the \emph{multiplexing gain} $r$ of the scheme according to
\begin{equation} \label{eq:multiplexing-gain}
r \defeq \lim_{\rho \rightarrow \infty} \frac{R(\rho)}{\log_{2} \rho} = \lim_{\rho \rightarrow \infty} \frac{1}{T} \frac{\log |\Xset(\rho)|}{\log \rho} \, .
\end{equation}
As we will be interested in the system behavior as a function of the multiplexing gain $r$, we will use the term \emph{design} to denote a set of schemes over some range of $r$. In this sense we would consider the Alamouti code \cite{Ala:98} or V-BLAST \cite{WFG:98} with appropriately chosen constellations as designs (c.f., \cite[Section VII]{ZT:03}). We will in what follows write $\Xset_{r}$ to express the dependence of the codebook (or more appropriately the sequence of codebooks) on the $r$, while the dependence on $\rho$ is suppressed for notational reasons. The \emph{diversity gain} of the design under ML decoding is given, as a function of $r$, according to (c.f.\ \cite{ZT:03})
\begin{equation} \label{eq:ml-diversity}
d_{\ml}(r) \defeq - \lim_{\rho \rightarrow \infty} \frac{\log \prob{\bfxh_{\ml} \neq \bfx}}{\log \rho}
\end{equation}
(provided the limit exists) where $\bfx$ is assumed uniformly distributed over $\Xset_{r}$ and where $\bfxh_{\ml}$ is given by \eqref{eq:ml} for $\Xset = \Xset_{r}$. The expression in \eqref{eq:ml-diversity} will in general define a tradeoff between the multiplexing gain and diversity gain, particular to the design and channel at hand \cite{ZT:03}.

As shown in \cite[Lemma 5]{ZT:03} the diversity gain $d_{\ml}(r)$ is under the power constraint, $\expt{\| \bfx \|^2} \leq T$, upper bounded by the outage exponent $d_{\out}(r)$ where
\begin{equation} \label{eq:outage-exponent}
d_{\out}(r) = -\lim_{\rho \rightarrow \infty} \frac{\log \mathrm{P} \big( \log \det(\bfI + \bfH\bfH\tr) \! < \! 2 R T \big) }{\log \rho} \, .
\end{equation}
In the case of the i.i.d.\ Rayleigh fading quasi-static MIMO channel (c.f., Section~\ref{sec:mimo-flat}), $d_{\out}(r)$ is given by the piece-wise linear curve connecting $(k,(\nr-k)(\nt-k))$ for $k=1,\ldots,\min(\nt,\nr)$ \cite{ZT:03}. Similar results have been obtained for other fading distributions \cite{ZMM:07}. A code is said to be approximately universal \cite{TV:06} for the particular system model at hand if $d_{\ml}(r) = d_{\out}(r)$ under any fading distribution. For the $\nt \times \nr$ quasi-static MIMO channel, approximately universal codes have been constructed for all $r$, $\nr$ and $\nt$ provided $T \geq \nt$ \cite{EKP:05,EKP:06}.

As frequently done in works on the DMT, we will make use of the $\doteq$ notation where $f(\rho) \doteq \rho^{x}$ iff (c.f., \cite{ZT:03})
\begin{equation} \label{eq:doteq}
\lim_{\rho \rightarrow \infty} \frac{\log f(\rho)}{\log \rho} = x \, .
\end{equation}
The symbols $\dotgeq$ and $\dotleq$ are defined similarly. In this notation a scheme has multiplexing gain $r$ if $|\Xset| \doteq \rho^{rT}$ and diversity gain $d$ under ML decoding if $\prob{\bfxh_{\ml} \neq \bfx} \doteq \rho^{-d}$.

\section{Lattice codes and decoding}

\subsection{Lattice designs} \label{sec:lattice-designs}

An $n$-dimensional real valued lattice $\Lambda$ is the discrete additive subgroup of $ \reals^n$ given by
\begin{equation} \label{eq:lattice}
\Lambda \defeq \{ \bfG \bfz \; |Ê\; \bfz \in \ints^n \} \, .
\end{equation}
The full rank matrix $\bfG \in \reals^{n \times n}$ is referred to as the generator matrix of $\Lambda$. We shall throughout consider a class of designs given as follows.

\vspace{3pt}
\begin{definition}[Lattice design]
A lattice design is defined by the pair $(\Lambda,\Rset)$ where $\Lambda \subset \reals^n$ is a lattice and $\Rset$ is a compact (i.e., closed and bounded) convex subset of $\reals^n$, which contains $\bfZero$ in its interior. For $r \geq 0$ the sequence of lattice codes $\Xset_{r}$ is given by $\Xset_{r} = \Lambda_{r} \cap \Rset$ where $\Lambda_{r} \defeq \phi_{r} \Lambda$ and $\phi_{r} \defeq \rho^{-\frac{rT}{n}}$.
\end{definition}
\vspace{3pt}

As in \cite{GCD:04}, we refer to $\Rset$ as the \emph{shaping region} of the lattice design. It is important to note that we assume that $\Rset$ and $\Lambda$ are fixed and independent of $\rho$ and that, in general, $\Rset$ has to be appropriately chosen so that the design satisfy the given power constraint, e.g., $\expt{\|Ê\bfx \|^2} \leq T$. This definition of a lattice design is slightly more restrictive than the definition of lattice space-time codes considered in \cite{GCD:04}, in that we require the same lattice (and shaping region) to be used for all multiplexing gains $r$ and SNR $\rho$. Note, however, that while we restrict the maximum value of $\|\bfx\|^2$ by the shaping region, we are not restricting the analysis to short-term power constraints, as long-term power allocation policies may often be considered part of the effective channel $\bfH$.

It is straightforward to verify that the multiplexing gain of $\Xset_{r}$ is indeed $r$. By a principle, dating back to Gauss, stating that the number of lattice points in a large set is well approximated by the volume of the set, we have \cite{BB:99}
\begin{align} \label{eq:ld-size}
|\Xset_{r}| = & |\Lambda_{r} \cap \Rset| = |\phi_{r} \Lambda \cap \Rset| \nonumber \\
= & |\Lambda \cap \phi_{r}^{-1} \Rset| = \frac{\phi_{r}^{-n} \vol(\Rset)}{\vol(\Vset_{\Lambda})} + o(\phi_{r}^{-n}) \doteq \rho^{rT}
\end{align}
where $\vol(\Rset)$ and $\vol(\Vset_{\Lambda})$ denotes the volume of the shaping region and the fundamental (Voronoi) cell of $\Lambda$ respectively.

The assumption that $\bfG$ is a square matrix can be made without loss of generality. To see this assume that $\bfG \in \reals^{n \times k}$ where $k < n$ and note that for $\bfx \in \Lambda$ we have $\bfx = \bfG \bfz$ for $\bfz \in \ints^k$. Write $\bfG = \bfU \bfG'$ where $\bfU \in \reals^{n \times k}$ has orthogonal columns and $\bfG' \in \reals^{k \times k}
$ is full rank, let $\bfH' = \bfH \bfU\tr$ and $\bfx' = \bfG' \bfz$. We obtain $\bfH \bfx = \bfH \bfG \bfz = \bfH' \bfU\tr \bfU \bfG' \bfz = \bfH' \bfG' \bfz = \bfH' \bfx'$, i.e., transmitting $\bfx$ over $\bfH$ is equivalent to transmitting $\bfx'$ over $\bfH'$. As no explicit assumption is made regarding the fading distribution of $\bfH$, we may equivalently consider the channel given by $\bfH'$, and use the square generator matrix $\bfG'$ in the formulation of the lattice design. The two equivalent cases naturally result in the same DMT curve. On the other hand, if $k > n$ we may extend $\bfG \in \reals^{n \times k}$ to a $k \times k$ full rank matrix by the addition of $k-n$ linearly independent rows, while adding $k-n$ columns containing zeros to $\bfH$ in the corresponding positions, thus leaving the input-output relation of \eqref{eq:model} unaltered.

The definition of a lattice design admits most of the codes mentioned in Section~\ref{sec:coding} in a straightforward manner in the sense that the code construction may be completely described by the pair ($\Lambda,\Rset)$. The largest subclass of lattice codes, generally known as linear dispersion codes  (c.f., \cite{HH:02} and \cite{GCD:04}), additionally satisfy $\bfx = \phi_{r} \sum_{i=1}^{n/2} (\bfa_{i} \alpha_{i} + \bfb_{i} \beta_{i})$ for some fixed $\bfa_{i},\bfb_{i} \in \reals^n$, $i=1,\ldots,n/2$, where $\alpha_{i}$ and $\beta_{i}$ constitute the real and imaginary part of a complex constituent data symbol chosen from a suitable constellation, e.g., a QAM or HEX \cite{FGL:04} constellation. The structure of the linear dispersion codes provides efficient encoding, and naturally yields a shaping region $\Rset$ in the form of an orthotope with axes aligned with the columns of the corresponding generator matrix $\bfG = [\bfa_{1}, \bfb_{1}, \ldots, \bfa_{n/2}, \bfb_{n/2}]$. The class of linear dispersion codes include the constructions in \cite{EKP:05,EKP:06,SRS:03,BR:03,KR:05,ORB:06,ESK:07,YB:07a,Lu:08,EVA:09,PKE:09,EK:09} as well as many classical designs \cite{Ala:98,TJC:99,WFG:98}. Also the codes with reduced decoding complexity in \cite{BHV:09,TK:02,PGG:07,SF:07,MS:06} belong to this class of codes. It is known that a better shaping gain may be achieved through a more careful design of the shaping region $\Rset$ (c.f., \cite{GCD:04,KC:09})

Before continuing, two remarks are in order. While \cite{KC:09} defines single lattices which provide strong lattice codes, the specific encoding strategy proposed in \cite{KC:09} will in general also introduce a (pseudo-random) translate of the lattice $\Lambda_{r}$. This is not covered by our basic definition of lattice designs which specifies the code exclusively in terms of $\Lambda$ and $\Rset$. Although the results presented in the following straightforwardly extend to cover such lattice translates, we shall in the interest of notational simplicity not consider this at first. Instead, we outline the changes required to handle this generalization in Section~\ref{sec:general}.  Furthermore we remark that we make no assumptions regarding the optimality of the code design itself, i.e., we do not assume that $d_{\ml}(r) = d_{\out}(r)$, and consequently the results are applicable also to suboptimal designs such as, e.g., V-BLAST.

\subsection{Lattice decoding} \label{sec:lattice-decoding}

The ML decoder in \eqref{eq:ml} implements a search for the codeword closest to $\bfy$ over $\Xset_{r} = \Lambda_{r} \cap \Rset$ \cite{AEV:02,DGC:03}. As in \cite{GCD:04,DGC:03} we use the term \emph{lattice decoding} to refer to an unconstrained search over $\Lambda_{r}$, i.e., a search where the constraint imposed by $\Rset$ is ignored by the decoder. The rationale behind such an approach is that it symmetrizes the problem and allows for the structure of the lattice to be exploited in order to reduce the computational complexity of the decoder \cite{AEV:02,DGC:03,MGD:06}.

The \emph{naive lattice decoder} (c.f.\ \cite{GCD:04}) is obtained by simply removing the constraint imposed by $\Rset$ in the ML decoder while keeping the decision metric unaltered, i.e.,
\begin{equation} \label{eq:naive}
\bfxh_{\nlat} = \arg \min_{\bfxh \in \Lambda_{r}}  \|Ê\bfy - \bfH \bfxh \|^2 \, .
\end{equation}
In the event that $\bfxh_{\nlat} \notin \Xset_{r}$ the decoder declares an error. It is known that the performance loss incurred by neglecting the codebook boundary $\Rset$ may in this case be substantial, and that the naive lattice decoder is not DMT optimal in general \cite{GCD:04,TK:07}. Still, as proved in \cite{GCD:04} for the i.i.d.\ Rayleigh fading quasi-static MIMO channel, the problem does not lie with lattice decoding per se, but with the naive implementation. In particular, after an appropriate alteration of the decoding metric, it was by a random coding argument shown that lattice coding and decoding is sufficient for achieving optimal DMT performance in this scenario \cite{GCD:04}.

Intuitively, as the naive lattice decoder \eqref{eq:naive} is suboptimal in terms of its diversity, it must mean that $\bfxh_{\nlat} \neq \Rset$ with a probability that is large in relation to $\prob{\bfxh_{\ml} \neq \bfx}$, i.e., the decoder is relatively likely to decide in favor of a codeword outside the region defined by $\Rset$. As $\Rset$ is bounded it is plausible that a regularization \cite{TGS:95} of the decoding metric may reduce the probability of ``out of region'' error events, and improve the probability of error.

\subsection{DMT optimality of regularized lattice decoding} \label{sec:main}

The (general) \emph{regularized lattice decoder} is given by
\begin{equation} \label{eq:lattice-decoder}
\bfxh_{\lat} = \arg \min_{\bfxh \in \Lambda_{r}}  \|Ê\bfy - \bfH \bfxh \|^2 + \|Ê\bfxh \|^2_{\bfT}
\end{equation}
where $\| \bfxh \|_{\bfT}^2 = \bfxh\tr \bfT \bfxh$ for some given positive definite matrix $\bfT = \bfT\tr$. The additive term $\|Ê\bfxh \|^2_{\bfT}$ applies an incremental penalization to lattice points further from the origin, and reduces the probability of error associated with codewords outside of the shaping region. This notion is formalized by the following theorem, which constitutes one of the main contributions of this work, and states that \eqref{eq:lattice-decoder} is a DMT optimal decoding strategy for lattice designs, in a remarkably general sense. The proof is given in Section~\ref{sec:proof}.

\vspace{3pt}
\begin{theorem} \label{thrm:main}
For any lattice design $(\Lambda,\Rset)$, and for any fading distribution such that $d_{\ml}(r)$ is (right) continuous at $r$, the regularized lattice decoder is DMT optimal, i.e.,
\begin{equation} \label{eq:main}
d_{\lat}(r) = d_{\ml}(r) \, ,
\end{equation}
where
\begin{equation} \label{eq:lat-diversity}
d_{\lat}(r) \defeq - \lim_{\rho \rightarrow \infty} \frac{\log \prob{\bfxh_{\lat} \neq \bfx}}{\log \rho} \, ,
\end{equation}
for $\bfx$ uniformly distributed over $\Xset_{r}$, and $\bfxh_{\lat}$ given by \eqref{eq:lattice-decoder}.
\end{theorem}
\vspace{3pt}

Before proving Theorem~\ref{thrm:main}, we remark that for $\bfT = \bfI$ the regularized decoder is equivalent to the MMSE-GDFE decoder considered in \cite{GCD:04}, if we neglect the lattice translate considered therein. In particular, the regularized lattice decoder in \eqref{eq:lattice-decoder} is equivalently given by (c.f., Appendix~\ref{app:equiv})
\begin{equation} \label{eq:mmse-gdfe}
\bfxh_{\lat} = \arg \min_{\bfxh \in \Lambda_{r}} \| \bfF \bfy -  \bfB \bfxhÊ\|^2
\end{equation}
where $\bfF \in \reals^{n \times m}$ and $\bfB \in \reals^{n \times n}$ are MMSE-GDFE forward and feedback filters \cite{GCD:04}. This equivalence is interesting in light of the fact that the motivation of the MMSE-GDFE decoder in \cite{GCD:04} was largely information theoretic in nature, while the regularization view is arguably of a more signal processing flavor. Theorem~\ref{thrm:main} thus extends the results of \cite{GCD:04} and proves DMT optimality of MMSE-GDFE decoding for any lattice designs based on a single, fixed, generator matrix. We also note that although the specific matrix $\bfT$ in \eqref{eq:lattice-decoder} has no effect on the diversity gain (provided $\bfT$ is full rank) it may significantly affect the coding gain and should in practice be chosen based on the shaping region, code, and channel statistics.

\subsection{Proof of Theorem~\ref{thrm:main}} \label{sec:proof}

We begin by providing the following lemma, proven in Appendix~\ref{app:lemma}. The purpose of the lemma is to connect the probably of ML error with the existence of a small codeword difference $\|Ê\bfH (\bfxh_{1}-\bfxh_{2})\|^2$ where $\bfxh_{1}$ and $\bfxh_{2}$ belong to a subset of the codebook. In essence, the lemma provides a ``deep fade typical error'' probability bound in line with \cite[Ch.\ 3]{TV:05}.

\vspace{3pt}
\begin{lemma} \label{lm:dmin}
Let $\Bset$ be the spherical region given by
\begin{equation} \label{eq:ball}
\Bset \defeq \{ \bfd \in \reals^n \, | \, \|Ê\bfd \|^2 \leq \gamma \}
\end{equation}
where the radius $\gamma > 0$ (independent of $\rho$) is chosen such that $\bfd_{1} + \bfd_{2} \in \Rset$ for any $\bfd_{1}, \bfd_{2} \in \Bset$. Let
\begin{equation} \label{eq:mindist}
\nu_{r} \defeq \min_{\bfd \in \Bset \cap \Lambda_{r} : \bfd \neq \bfZero} \tfrac{1}{4} \|Ê\bfH \bfd \|^2 \, .
\end{equation}
Then, for any $r > 0$ it holds that
\begin{equation} \label{eq:outage}
\limsup_{\rho \rightarrow \infty} \frac{\log \prob{\nu_{r} \leq 1}}{\log \rho} \leq -d_{\ml}(r) \, .
\end{equation}
\end{lemma}
\vspace{3pt}

The existence of the set $\Bset$ in \eqref{eq:ball} follows by the assumption that $\bfZero$ is contained in the interior of $\Rset$. Now, let $\zeta > 0$ be given and choose $\delta > 0$ such that
\begin{equation} \label{eq:delta}
\frac{2 \zeta T}{n} > \delta > 0 \, .
\end{equation}
This may clearly be done for arbitrary $\zeta > 0$. We will in the following assume that $\nu_{r + \zeta} \geq 1$ and that $\| \bfw \|^2 \leq \rho^{\delta}$, and prove that these two conditions are sufficient for a correct decision by the regularized lattice decoder in \eqref{eq:lattice-decoder}, provided that $\rho$ is sufficiently large. Hence, in order for an error to occur at large $\rho$, one of the assumptions must fail.

To this end, consider first the metric in \eqref{eq:lattice-decoder} for the transmitted codeword $\bfx$, i.e.,
\begin{equation} \label{eq:metric-transmitted}
\| \bfy - \bfH\bfx \|^2 + \| \bfx \|^2_{\bfT} \leq \rho^\delta + c
\end{equation}
where $\bfy - \bfH \bfx = \bfw$ and $\| \bfw \|^2 \leq \rho^{\delta}$ was used, and where
$$
c \defeq \max_{\bfr \in \Rset} \| \bfr \|^2_{\bfT} \, .
$$
Note that $c < \infty$ as $\Rset$ is bounded and that $c$ is independent of the transmitted codeword $\bfx$ and $\rho$.

In order to bound the metric for $\bfxh \in \Lambda_{r}$ where $\bfxh \neq \bfx$, we note that $\nu_{r + \zeta} \geq 1$ implies
\begin{equation} \label{eq:bound1}
\tfrac{1}{4} \|Ê\bfH \bfd \|^2 \geq 1 \quad \forall \, \bfd \in \Bset \cap \Lambda_{r+\zeta} \, , \, \bfd \neq \bfZero \, ,
\end{equation}
by the definition in \eqref{eq:mindist}. As $\Lambda_{r} = \rho^{\frac{\zeta T}{n}} \Lambda_{r+\zeta}$ it follows that
\begin{equation} \label{eq:bound2}
\tfrac{1}{4} \|Ê\bfH \bfd \|^2 \geq \rho^{\frac{2 \zeta T}{n}} \quad \forall \, \bfd \in \rho^{\frac{\zeta T}{n}}\Bset \cap \Lambda_{r} \, , \, \bfd \neq \bfZero
\end{equation}
after scaling \eqref{eq:bound1} by $\rho^{\frac{\zeta T}{n}}$. As $\Rset$ is bounded, and as $\zeta > 0$, it holds that $\Rset \subset \tfrac{1}{2}\rho^{\frac{\zeta T}{n}} \Bset$ for all $\rho \geq \rho_{1}$, given some sufficiently large $\rho_{1}$. This implies that $\bfx \in \tfrac{1}{2}\rho^{\frac{\zeta T}{n}} \Bset$ for $\rho \geq \rho_{1}$ since $\bfx \in \Rset$. It is important to note here that while $\rho_{1}$ may depend on $\zeta$ and $\Rset$, it can be chosen independent of the particular $\bfx$ transmitted.

For any $\bfxh \in \tfrac{1}{2}\rho^{\frac{\zeta T}{n}} \Bset \cap \Lambda_{r}$, $\bfxh \neq \bfx$, it holds that $\bfd = \bfx - \bfxh \in \rho^{\frac{\zeta T}{n}} \Bset \cap \Lambda_{r}$. By \eqref{eq:bound2} we have
\begin{equation} \label{eq:bound3}
\tfrac{1}{4} \| \bfH (\bfx - \bfxh) \|^2 = \tfrac{1}{4} \| \bfH \bfd \|^2 \geq \rho^{\frac{2 \zeta T}{n}}
\end{equation}
where $\bfd = \bfx - \bfxh$. As $\| \bfw \|^2 \leq \rho^\delta$ it follows by \eqref{eq:bound3} and \eqref{eq:delta} that $\tfrac{1}{4} \| \bfH \bfd\|^2 \gg \|Ê\bfw \|^2$ for large $\rho$. In particular, there is some $\rho_{2} \geq \rho_{1}$, independent of $\bfx$ and $\bfxh$, for which the triangle inequality implies that
$$
\| \bfy - \bfH \bfxh \|^2 = \|Ê\bfH(\bfx-\bfxh) + \bfw \|^2 \geq \rho^{\frac{2 \zeta T}{n}}
$$
for all $\rho \geq \rho_{2}$. Consequently,
\begin{equation} \label{eq:metric-inball}
\| \bfy - \bfH \bfxh \|^2 + \|Ê\bfxh \|^2_{\bfT} \geq \rho^{\frac{2 \zeta T}{n}}
\end{equation}
for any $\bfxh \in \Lambda_{r}$ where $\bfxh \in \tfrac{1}{2} \rho^{\frac{\zeta T}{n}}\Bset$ and $\rho \geq \rho_{2}$.

In the case that $\bfxh \notin \tfrac{1}{2} \rho^{\frac{\zeta T}{n}}\Bset$, it follows by the definition in \eqref{eq:ball} that $\| \bfxh \|^2 \geq \frac{1}{4} \gamma \rho^{\frac{2 \zeta T}{n}}$ which implies $\| \bfxh \|^2_\bfT \geq \frac{1}{4} \gamma \lambda_{\min}(\bfT) \rho^{\frac{2 \zeta T}{n}}$ where $\lambda_{\min}(\bfT) > 0$ denotes the minimum eigenvalue of $\bfT$. It follows that
\begin{equation} \label{eq:metric-outball}
\| \bfy - \bfH \bfxh \|^2 + \|Ê\bfxh \|^2_{\bfT} \geq \tfrac{1}{4} \gamma \lambda_{\min}(\bfT) \rho^{\frac{2 \zeta T}{n}}
\end{equation}
for any $\bfxh \notin \rho^{\frac{\zeta T}{n}}\Bset$.

Let
\begin{equation} \label{eq:bounds}
a(\rho) \defeq \rho^\delta + c \quad \text{and} \quad b(\rho) \defeq \min \! \big( 1 ,\tfrac{1}{4}\gamma \lambda_{\min}(\bfT) \big) \rho^{\frac{2 \zeta T}{n}}
\end{equation}
and note that \eqref{eq:delta} implies that there is some $\rho_{3} \geq \rho_{2}$, again independent of $\bfx$ and $\bfxh$, for which $a(\rho) < b(\rho)$ for all $\rho > \rho_{3}$. For the transmitted codeword $\bfx$ we have by \eqref{eq:metric-transmitted} that
$$
\| \bfy - \bfH \bfx \|^2 + \|Ê\bfx \|^2_{\bfT} \leq a(\rho) \, .
$$
For any other $\bfxh \in \Lambda_{r}$ (i.e., $\bfxh \in \Lambda_{r} \backslash \{Ê\bfx \}$) it holds by \eqref{eq:metric-inball} and \eqref{eq:metric-outball} that
\begin{equation} \label{eq:tight-bound}
\| \bfy - \bfH \bfxh \|^2 + \|Ê\bfxh \|^2_{\bfT} \geq b(\rho) > a(\rho)
\end{equation}
for all $\rho \geq \rho_{3}$. This implies that the transmitted codeword yields the minimum metric in \eqref{eq:lattice-decoder}, or equivalently that $\bfxh_{\lat} = \bfx$ as long as $\rho \geq \rho_{3}$ and under the assumptions that $\nu_{r+\zeta} \geq 1$ and $\| \bfw \|^2 \leq \rho^{\delta}$. For an error to occur when $\rho \geq \rho_{3}$ it is thus required that $\nu_{r+\zeta} < 1$ or $\|\bfw\| > \rho^{\delta}$.

Applying the union bound to the probability of error yields
\begin{equation} \label{eq:union-bound}
\prob{\bfxh_{\lat} \neq \bfx} \leq \prob{\nu_{r+\zeta} < 1} + \prob{\|\bfw\| > \rho^{\delta}} \, ,
\end{equation}
for $\rho \geq \rho_{3}$. As $\prob{\|\bfw\| > \rho^{\delta}} \doteq \rho^{-\infty}$, due to the exponential tail of the Gaussian distribution, the second term in \eqref{eq:union-bound} is asymptotically irrelevant. By Lemma~\ref{lm:dmin} it follows that $\prob{\nu_{r+\zeta} < 1} \dotleq \rho^{-d_{\ml}(r+\zeta)}$. Note here also that Lemma~\ref{lm:dmin} is applicable even when $r = 0$ since it is applied at a multiplexing gain of $r + \zeta > 0$. It follows that
\begin{equation} \label{eq:last-limit}
\limsup_{\rho \rightarrow \infty} \frac{\log \prob{\bfxh_{\lat} \neq \bfx}}{\log \rho} \leq -d_{\ml}(r+\zeta) \, .
\end{equation}
By observing that \eqref{eq:last-limit} holds for an arbitrary choice of $\zeta > 0$, we may conclude that
\begin{equation} \label{eq:last-limit2}
\limsup_{\rho \rightarrow \infty} \frac{\log \prob{\bfxh_{\lat} \neq \bfx}}{\log \rho} \leq -d_{\ml}(r) \, .
\end{equation}
for any $r \geq 0$, provided that
$$
\lim_{\zeta \rightarrow 0^{+}} d_{\ml}(r+\zeta) = d_{\ml}(r) \, ,
$$
i.e., provided $d_{\ml}(r)$ is right continuous at $r$. As $\prob{\bfxh_{\lat} \neq \bfx} \geq \prob{\bfxh_{\ml} \neq \bfx}$ due to the optimality of the ML decoder it holds that
$$
\liminf_{\rho \rightarrow \infty} \frac{\log \prob{\bfxh_{\lat} \neq \bfx}}{\log \rho} \geq -d_{\ml}(r) \, .
$$
which combined with \eqref{eq:last-limit2} establish the claim of Theorem~\ref{thrm:main}.

\subsection{A geometric example}

\begin{figure*}

\psfrag{R}[cc]{$\Rset$}
\psfrag{xh}[cc]{$\bfxh$}
\psfrag{x}[cc]{$\bfx$}
\psfrag{y}[cc]{$\bfy$}
\psfrag{yh}[cc]{$\bfyh$}
\psfrag{z}[cc]{$\bfr$}
\psfrag{zh}[cc]{$\bfrh$}
\psfrag{v1}[cc]{$\bfv_{1}$}
\psfrag{v2}[cc]{$\bfv_{2}$}
\psfrag{u1}[cc]{$\bfu_{1}$}
\psfrag{u2}[cc]{$\bfu_{2}$}

\centering
\subfigure[Original lattice $\Lambda_{r}$ and shaping region $\Rset$]{
\includegraphics[width=5.62cm]{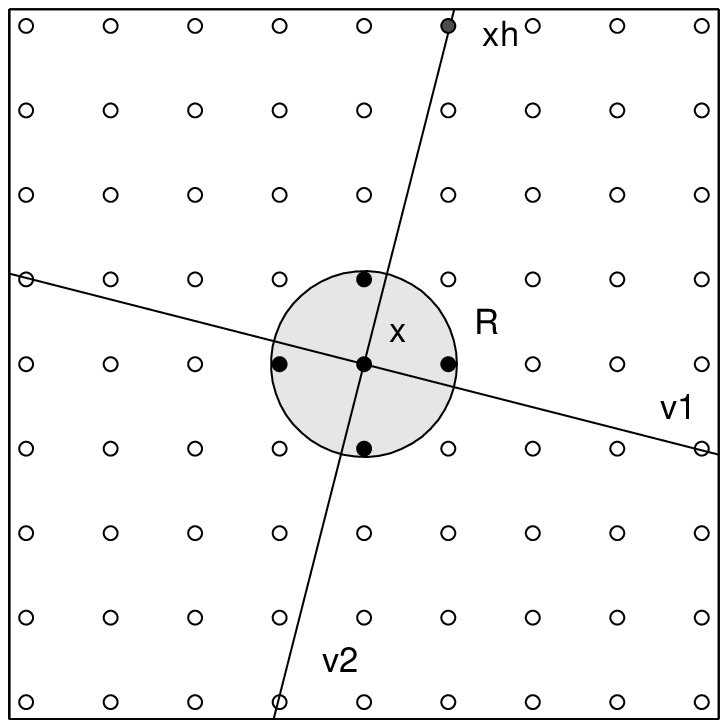}
\label{fig:lattice}
}
\subfigure[Image of $\Lambda_{r}$ and $\Rset$ under linear map $\bfH$]{
\includegraphics[width=5.71cm]{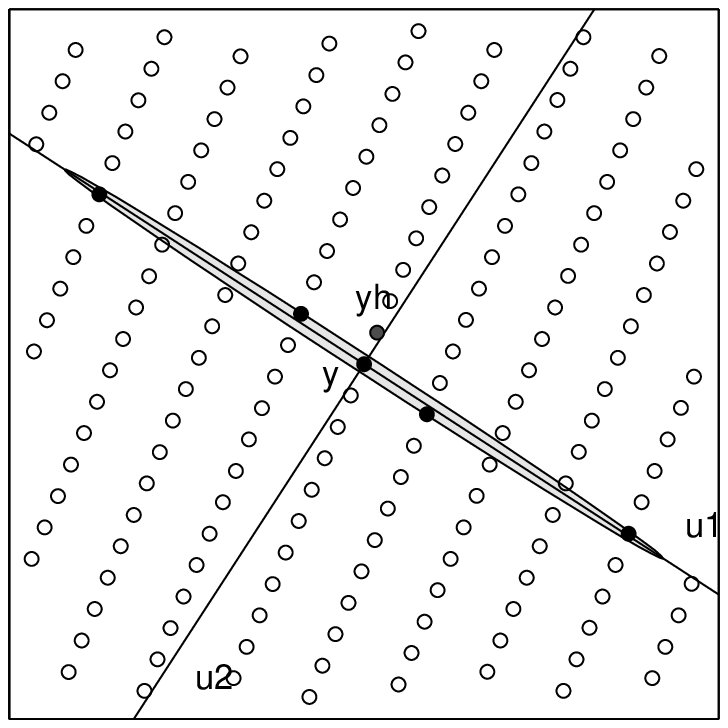}
\label{fig:naive}
}
\subfigure[Image of $\Lambda_{r}$ and $\Rset$ under linear map $\bfB$]{
\includegraphics[width=5.68cm]{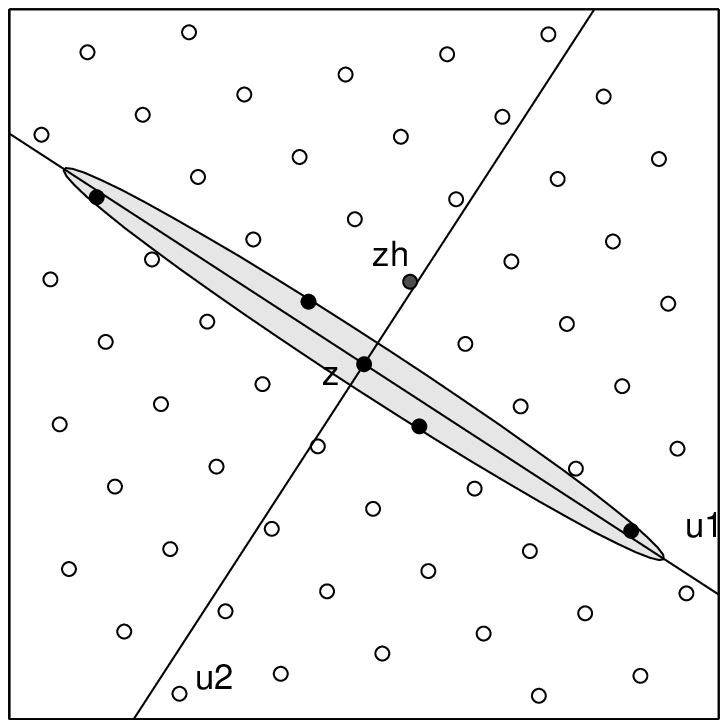}
\label{fig:mmse}
}
\label{fig:subfigureExample}
\caption{Transformation of lattice $\Lambda_{r}$ and spherical shaping region $\Rset$ under linear map induced by channel $\bfH$ and MMSE filter $\bfB$. Singular vectors of $\bfH = \bfU \bfSigma \bfV\tr$, where $\bfU = (\bfu_{1},\bfu_{2})$ and $\bfV = (\bfv_{1},\bfv_{2})$, are shown as solid lines for reference. The matrix $\bfB$ is such that $\bfB\tr\bfB = \bfI + \bfH\tr\bfH$ where $\bfB$ shares left and right singular vectors with $\bfH$. Further, $\bfy = \bfH \bfx$, $\bfyh = \bfH \bfxh$, $\bfr = \bfB\bfx$, and $\bfrh = \bfB \bfxh$.
} \label{fig:mapping}
\end{figure*}

In order to provide further intuition into the suboptimality of the naive lattice decoder, and the argument made in Section~\ref{sec:proof} it is useful to consider the example provided in Fig.\ \ref{fig:mapping}, where $\Lambda_{r}$ is a scaled version of the integer lattice $\ints^2$ and where the shaping region $\Rset$ is spherical. The image of $\Lambda_{r}$ and $\Rset$ under the linear map induced by $\bfH$ are shown in Fig.\ \ref{fig:naive}. In the example, $\bfH \in \reals^{2 \times 2}$ is nearly rank deficient. For the illustration, $\sigma_{1}(\bfH) = 40 \sigma_{2}(\bfH)$ where $\sigma_{i}(\bfH)$ denotes the $i$th singular value of $\bfH$.

We will in the following discussion assume that $\bfx = \bfZero$ corresponds to the transmitted codeword and, for simplicity, that $\bfT = \bfI$. As seen in Fig.\ \ref{fig:naive} no other codeword $\bfxh \in \Xset_{r} \backslash \{\bfx\}$ is mapped close to $\bfH\bfx$ by the linear map $\bfH$. Thus, the ML decoder is unlikely to make an error. However, when considering decoding to the full lattice $\Lambda_{r}$, the (naive) lattice decoder is likely to decide in favor of the, in Fig.\ \ref{fig:lattice}, indicated codeword $\bfxh \in \Lambda_{r}$. This is a consequence of the fact that $\bfxh$ lies close to the space spanned by the right singular vector corresponding to the smallest singular value of $\bfH$ (c.f.\ Fig.\ \ref{fig:lattice}). The closeness of $\bfH \bfxh$ to $\bfH \bfx$ illustrates the problem with the naive lattice decoder, i.e., even when no codewords in $\Xset_{r}$ lie close to the space corresponding to a weak singular value of $\bfH$ it may be likely that a ``hypothetical'' codeword in $\Lambda_{r}$ does. This view is strengthened by the observation that the performance of the naive lattice decoder is often determined by the statistics of the channel's weakest eigenmode (c.f.\ \cite{TK:07,GCD:04}), although the fixed-rate V-BLAST result in \cite{TMK:07} provides an exception to this rule.

The intuitive argument behind the regularization is that any lattice point $\bfxh$ (far) outside the constellation region $\Rset$, which implies that $\|\bfxh\|^2$ is large, is significantly penalized by the regularized decision metric. For codewords $\bfxh \neq \bfx$ in $\Rset$ the first quadratic term in \eqref{eq:lattice-decoder} will be large, unless the ML decoder is also likely to be in error. Although this heuristic argument fails for codewords $\bfxh$ close to the boundary of $\Rset$, this problem may be circumvented under the continuity assumption of Theorem~\ref{thrm:main} by considering a larger constellation region, corresponding to the codebooks used at a marginally higher multiplexing gain.

The effect of the regularization can also be seen in Fig.\ \ref{fig:mmse} that shows the image of $\Lambda_{r}$ under the linear transformation of the MMSE-GDFE feedback filter $\bfB$ in \eqref{eq:mmse-gdfe}, corresponding to a regularized version of $\bfH$. For the purpose of the illustration, we have chosen $\bfB$ so that is shares left and right singular vectors with $\bfH$. While the image of codewords inside $\Rset$ under the transformations $\bfH$ and $\bfB$ are relatively similar (c.f., Fig. \ref{fig:naive} and \ref{fig:mmse}), codewords outside the constellation $\Rset$ are more affected by the change from $\bfH$ to $\bfB$. Note in particular the difference between $\bfyh = \bfH\bfxh$ and $\bfrh = \bfB\bfxh$ in Fig. \ref{fig:naive} and \ref{fig:mmse}. Decoding to the closest lattice point in Fig.\ \ref{fig:mmse} is in this case clearly a better approximation of the ML decoder than decoding to the closest lattice point in Fig.\ \ref{fig:naive}.

\section{Computationally Efficient Decoding} \label{sec:eff-dec}

\subsection{DMT optimality of approximate lattice decoding} \label{sec:c-approx}

Obtaining $\bfxh_{\lat}$ in \eqref{eq:lattice-decoder} still requires the minimization of a quadratic function over the discrete lattice $\Lambda_{r}$, a problem which is NP-hard in general, even after pre-processing \cite{Mic:01}. This implies that even if lattice reduction techniques are used when obtaining the exact solution to \eqref{eq:lattice-decoder}, it is unlikely that there will be any general techniques with a (worst-case) complexity that grows sub-exponentially in the problem dimension $n$, unless the code itself provides a structure that simplifies decoding, such as for example in the case of orthogonal designs \cite{Ala:98,TJC:99}. For most high-performance lattice codes no such efficient solutions to \eqref{eq:lattice-decoder} are known, which motivates the study of suboptimal implementations of the regularized lattice decoder.

The codeword $\bfxh_{\lat}$ is by definition the codeword which provides the minimum metric in \eqref{eq:lattice-decoder}. A \emph{$C$-approximate solution} to \eqref{eq:lattice-decoder} is any $\bfxh \in \Lambda_{r}$ which for $C > 1$ satisfies
\begin{equation} \label{eq:approx}
\xi(\bfxh) \leq C \xi(\bfxh_{\lat}) \quad \text{where} \quad \xi(\bfxh) = \|Ê\bfy - \bfH \bfxh \|^2 + \| \bfxh \|^2_{\bfT} \, .
\end{equation}
An algorithm that for fixed $C$ is capable of producing a $C$-approximate solution to \eqref{eq:lattice-decoder}, for arbitrary inputs $\bfy \in \reals^{m}$ and $\bfH \in \reals^{m \times n}$, is referred to as a \emph{$C$-approximation algorithm} \cite{Hro:02}. In what follows we prove that any $C$-approximation algorithm for \eqref{eq:lattice-decoder} is sufficient for DMT optimal decoding in the sense of Theorem~\ref{thrm:main}.

\vspace{3pt}
\begin{theorem}\label{thrm:approx}
For any lattice design $(\Lambda,\Rset)$, and fading distribution such that $d_{\ml}(r)$ is (right) continuous at $r$, all $C$-approximate implementations of the regularized lattice decoder are DMT optimal provided $C$ is independent of $\rho$, i.e.,
\begin{equation} \label{eq:approx-statement}
d_{\app}(r) = d_{\ml}(r) \, ,
\end{equation}
where
\begin{equation} \label{eq:app-diversity}
d_{\app}(r) \defeq - \lim_{\rho \rightarrow \infty} \frac{\log \prob{\bfxh_{\app} \neq \bfx}}{\log \rho} \, ,
\end{equation}
for $\bfx$ uniformly distributed over $\Xset_{r}$, and where $\bfxh_{\app}$ is any $C$-approximate solution to \eqref{eq:lattice-decoder}.
\end{theorem}
\vspace{3pt}

\noindent \emph{Proof:} The proof follows from the proof of Theorem~\ref{thrm:main}, provided in Section~\ref{sec:proof}. In particular, consider $a(\rho)$ and $b(\rho)$ defined in \eqref{eq:bounds}. By the assumption in \eqref{eq:delta} it follows that
$$
\lim_{\rho \rightarrow \infty} \frac{b(\rho)}{a(\rho)} = \infty \, .
$$
We may thus select $\rho_{4} \geq \rho_{3}$ such that $b(\rho) \geq C a(\rho)$ for all $\rho \geq \rho_{4}$. As the metric for the transmitted codeword $\bfx$ is upper bounded by $a(\rho)$, and the metric of any other codeword is lower bounded by $b(\rho)$, it follows that when $\rho \geq \rho_{4}$, the only $C$-approximate solution to \eqref{eq:lattice-decoder} is $\bfx$, i.e., $\bfxh_{\app} = \bfx$ for $\rho \geq \rho_{4}$, under the assumptions that $\nu_{r+\zeta} \leq 1$ and $\| \bfw \|^2 \leq \rho^\delta$. The remaining proof is then analogous to the proof of Theorem~\ref{thrm:main} in Section~\ref{sec:proof}. \hfill $\square$
\vspace{3pt}

\subsection{DMT optimality of LR-aided lattice decoding}

The existence of computationally efficient $C$-approximate solutions is thus of interest for DMT optimal decoding of lattice designs. Fortunately, such solutions are already known, both with respect to \eqref{eq:lattice-decoder}, or to the equivalent MMSE-GDFE formulation in \eqref{eq:mmse-gdfe}. In fact, as shown in Appendix~\ref{app:equiv}, any $C$-approximate solution to \eqref{eq:mmse-gdfe} is also a $C$-approximate solution to \eqref{eq:lattice-decoder}. Of special interest in the communications context is Babai's nearest plane algorithm \cite{Bai:86}, which is equivalent to the LLL-based \cite{LLL:82} LR-aided SIC solution to \eqref{eq:mmse-gdfe} \cite{YW:02, WF:03, Bai:86, MGD:06}. The nearest plane algorithm provides a computationally efficient $C_{1}$-approximate solution \eqref{eq:mmse-gdfe} with $C_{1} \defeq 2^{\frac{n}{2}}$ \cite{Bai:86}. Similarly, the LLL-based LR-aided linear solution to \eqref{eq:mmse-gdfe}, discussed in \cite{Bai:86} as the rounding algorithm, provides a $C_{2}$-approximate solution whith $C_{2} \defeq 1+2n(9/2)^{\frac{n}{2}}$. For completeness, we give the following corollary to Theorem~\ref{thrm:approx}.

\vspace{3pt}
\begin{corollary} \label{cor:lr-aided}
The efficient LLL-based LR-aided linear (or SIC) implementations of the regularized lattice decoders provide DMT optimal decoding of any lattice design under the assumptions made in Theorem \ref{thrm:main} and \ref{thrm:approx}.
\end{corollary}
\vspace{3pt}

\noindent \emph{Proof:} The corollary follows by the equivalence of the LR-aided linear decoder and the rounding algorithm in \cite{Bai:86}, or of the LR-aided SIC decoder and the nearest plane algorithm in \cite{Bai:86}, in conjunction with Theorem~\ref{thrm:approx}. \hfill $\square$

\vspace{3pt}

Corollary \ref{cor:lr-aided} applies directly to the LR-aided linear implementation of the MMSE-GDFE decoder \cite{DGC:03,DGC:04,MGD:06},
due to the equivalence of the MMSE-GDFE decoder and the regularized decoder as outlined in Appendix \ref{app:equiv}. The corollary applies also to the LR-aided MMSE-SIC decoder proposed in \cite{WBK:04}, when applied to the equivalent channel
$$
\bfy = \bfH \bfG \bfs + \bfw
$$
where $\bfs \in \phi_{r} \ints^n$. Note however that in the latter case we would have $\bfT = (\bfG \tr \bfG)^{-1}$, as opposed to $\bfT = \bfI$, reflecting a regularization of $\bfs$ rather than $\bfx = \bfG \bfs$. In the case of perfect codes \cite{ORB:06}, where $\bfG = \bfI$, the metric of the MMSE-GDFE and the MMSE-SIC decoder coincides.

Corollary \ref{cor:lr-aided} applies also to a time-limited implementation of the Schnorr-Euchner (SE) sphere decoder \cite{AEV:02,SE:94} operating in the LLL reduced regularized lattice, provided the sphere decoder tree-search is allowed to reach the first leaf-node. This follows as the first leaf-node found by the SE SD corresponds to the Babai-point, i.e., the solution obtained by the nearest plane algorithm (c.f., \cite{AEV:02}). Finding further candidate codewords with smaller metric can only improve the approximation ratio.

\subsection{Decoding complexity} \label{sec:lll}

Both the LR-aided SIC and linear decoders discussed above begin by LLL reducing the lattice generated by $\bfM = \bfB \bfG$, where $\bfG$ is the generator matrix of $\Lambda$ and where $\bfB$ is the MMSE feedback filter (c.f., \cite{GCD:04} and Appendix \ref{app:equiv}), followed by a SIC or linear decoding stage in the reduced basis. Note here that by the regularization of $\bfB$ the matrix $\bfM$ is always full rank which makes the LLL algorithm applicable, regardless of the channel realization and the system dimensionality. The complexity of the decoding stage is only $\Oset(n^2)$ \cite{YW:02,WF:03,WBK:04} while the pre-processing relying on the LLL reduction is more complex. It is therefore relevant to consider the complexity of the LLL algorithm when applied to $\bfM$ in order to address the complexity of DMT optimal decoding of lattice designs. We refer the reader to \cite{YW:02,WF:03,WBK:04} for the implementation details of LR-aided decoders.

The LLL algorithm provides an iterative approach to lattice reduction \cite{LLL:82}. The number $K$ of LLL iterations required to reduce a given lattice generator matrix $\bfM \in \reals^{n \times n}$ may be bounded according to \cite{JSM:08,DV:94}
\begin{equation} \label{eq:LLL-bound}
K \leq n^2 \log_{s} \kappa(\bfM) + n
\end{equation}
where $s = 2 / \sqrt{3}$ and where $\kappa(\bfM)$ denotes the 2-norm condition number of $\bfM$. Each iteration requires $O(n^2)$ floating point operations \cite{LLL:82}. The number of operations per iteration may, however, be reduced to $O(n)$ if only an effectively LLL-reduced basis is required, as is the case when a SIC decoder is applied in the reduced basis \cite{LH:07}.

It is important here to note that for arbitrary $\bfM \in \reals^{n \times n}$ there is no universal upper bound on the number of iterations required to reduce $\bfM$ \cite{JSM:08}. Thus, the worst-case complexity of the LLL-based LR-aided decoder is unbounded if applied to arbitrary channels. However, in order to achieve DMT optimal performance it is not required to LLL reduce every conceivable channel. To see this, consider a decoder implementation which is allowed to time-out, and declare an error, when the number of floating point operations exceeds a given threshold. Denote the time-out event $\Tset$, and note that as long as $\prob{\Tset} \dotleq \rho^{-d_{\ml}(r)}$ the time-limitation imposed will not reduce the diversity gain, or potential DMT optimality, of the decoder. In light of \eqref{eq:LLL-bound} we may thus limit the application of the LLL algorithm to bases $\bfM = \bfB \bfG$ with bounded condition number $\kappa(\bfM)$, or allow the decoder the option to time out, stop, and declare an error. In order to be able to provide an effective statement regarding the worst case decoding complexity under time-outs, we impose here a moderate restriction on the channels considered.

We say that a channel is \emph{power limited} if $\expt{\| \bfH \|^2\fro} \dotleq \rho$ and note that this is required whenever we wish to interpret the parameter $\rho$ as an average SNR at the receiver.  For the class of power limited channels we may make the following statement, proven in Appendix~\ref{app:power}.

\vspace{3pt}
\begin{lemma} \label{lm:power}
For any power limited channel there is some constant $\alpha > 0$ where for $\bfM = \bfB \bfG$ it holds that
\begin{equation} \label{eq:cond-bnd}
\prob{\kappa(\bfM) \geq \rho^\alpha} \dotleq \rho^{-d_{\ml}(r)} \, ,
\end{equation}
provided $d_{\ml}(r) < \infty$.
\end{lemma}
\vspace{3pt}

By applying Lemma~\ref{lm:power}, \eqref{eq:LLL-bound} and Corollary~\ref{cor:lr-aided}, together with the previous discussion, the following statement regarding the complexity of DMT optimal decoding can thus be made. Note here that the signal space dimension $n$ is considered fixed and is thus hidden in the big-$O$ expression.

\vspace{3pt}
\begin{theorem} \label{thrm:complex}
For power limited channels, over any range of multiplexing gains $r$ where $d_{\ml}(r)$ is continuous, DMT optimal decoding of any lattice design is feasible at a worst-case complexity of $O(\log \rho)$.
\end{theorem}
\vspace{3pt}

\noindent \emph{Proof:} The theorem follows by imposing the constraint $\kappa(\bfM) \leq \rho^\alpha$ in \eqref{eq:LLL-bound}, where $\alpha$ is chosen according to Lemma~\ref{lm:power}, and noting that such a restriction in the set of channels to which the decoder is applied does not reduce the diversity. \hfill $\square$\vspace{3pt}

Although the bound in Theorem~\ref{thrm:complex} implies an increase in the LLL LR complexity for increasing SNR, this complexity only grows linearly in $\log \rho$. By comparing to \eqref{eq:rate} and \eqref{eq:ld-size} it may be seen that this corresponds to a linear increase in complexity as a function of the rate $R$ at high SNR. The LLL complexity should also be put in context with the full search implementation of the ML decoder whose complexity is $|\Xset_{r}|$ and thus exponential in $R$. This also applies to sphere decoding implementations where the worst-case complexity reported (see for example \cite{BHV:09} for fast decodable codes \cite{TK:02,PGG:07,SF:07}) is also exponential in $R$, albeit with a smaller exponent than the full search.
The same holds true for the hybrid transceiver in \cite{MS:06} (given $\nt \geq 2$). All such lattice-based designs may, however, be DMT optimally decoded using an LR-aided regularized lattice decoder structure with $O(\log \rho)$ complexity, potentially at some loss in coding gain, but at no diversity loss.

Finally, we note that in the case where $d_{\lat}(r) = \infty$ the statement in \eqref{eq:cond-bnd} in Lemma~\ref{lm:power} cannot be guaranteed based on the condition that $\expt{\| \bfH \|^2} \dotleq \rho$ alone. However, for any channel statistics under which $\prob{\|\bfH\|^2Ê\geq \rho^\alpha} \doteq \rho^{-\infty}$ for some sufficiently large $\alpha$, Theorem~\ref{thrm:complex} still applies. This includes for instance the quasi-static MIMO channel (c.f., Section~\ref{sec:mimo-flat}) under i.i.d. Rayleigh fading, or any other fading distribution with exponential tails.

\subsection{The search for improved approximation algorithms}

It is in the context of $C$-approximation algorithms important to note that while DMT optimality follows for any finite $C$, the gap in terms of SNR to the optimal implementation of \eqref{eq:lattice-decoder} will in general depend on $C$. Thus, the loss in performance at practical SNR may be unacceptable for unduly large values of $C$. This motivates further study into new approximation algorithms, and code designs, that jointly yield improved approximation ratios.

Such methods may include stronger LR methods such as the \emph{deep insertion} LLL variant \cite{SE:94} that is more computationally expensive but which finds better bases. Other LR approaches include methods based on the Korkine-Zolotareff bases (c.f., \cite{AEV:02}), and the \emph{algebraic lattice reduction} approach in \cite{LOB:08}. The latter method was presented for the $2\times 2$ golden code \cite{BRV:05} over the quasi-static MIMO channel, and approximates the channel matrix with the matrix representation of an invertible element of the maximal order of the CDA. Codes in which the ML decoder may be applied to spaces of reduced dimensionality (c.f., \cite{KR:09,HR:07}, as well as \cite{BHV:09,TK:02,PGG:07,SF:07}) may benefit from a reduced gap between ML and lattice decoding due to the general dependence of the approximation constant $C$ and the lattice dimension. This would suggest the use of transceivers based on reduced-dimensionality codes and regularized lattice decoding, as a good way to further approach ML error performance with a reduced SNR penalty. The topic of $C$-approximate solutions is, however, in the context of space-time decoding relatively unexplored at this stage.

\section{Generalizations} \label{sec:general}

In this section we consider a few straightforward generalizations in terms of the class of designs covered by the results as well as the modeling assumptions imposed in Section~\ref{sec:model}.

\subsection{Nested lattice designs} \label{sec:shaping} In the proof of Theorem~\ref{thrm:main}, and in the lattice designs of Section~\ref{sec:lattice-designs}, we assume a fixed shaping region $\Rset$, applied for all $\rho$. This condition could, however, be relaxed in favor of a sequence of shaping regions $\Rset(\rho)$, such that $\underline{\Rset} \subset \Rset(\rho) \subset \overline{\Rset}$ for sufficiently large $\rho$ where $\underline{\Rset}$ and $\overline{\Rset}$ are fixed ``inner'' and ``outer'' shaping regions that satisfy the conditions in Section~\ref{sec:lattice-designs}. Such an extension could be of interest for nested lattice codes \cite{EZ:04} involving a shaping lattice $\underline{\Lambda}_{r}$ satisfying $\underline{\Lambda}_{r} \subset \Lambda_{r}$ where $\Rset$ is the Voronoi region of $\underline{\Lambda}_{r}$, i.e., $\Rset = \Vset_{\underline{\Lambda}_{r}}$ \cite{EZ:04,GCD:04,KC:09}. One option along this line is to let $\underline{\Lambda}_{r} = \omega_{r} \Lambda_{r}$ where $\omega_{r} \in \naturals$ is an appropriately selected integer (i.e., self-similar nesting \cite{GCD:04}). This will in general require $\Rset$ to weakly depend on $\rho$, if we wish the code to be properly defined for all $r$ and $\rho$. Alternatively, self-similar nested designs could also be accommodated by replacing the assumption that $\phi_{r} = \rho^{\frac{-rT}{n}}$ by the relaxed assumption $\phi_{r} \doteq \rho^{-\frac{rT}{n}}$, e.g., $\phi_{r}^{-1} = \lceil \rho^{\frac{rT}{n}} \rfloor$ where $\lceil \cdot \rfloor$ denotes rounding to the nearest integer. The proof given in Section~\ref{sec:proof} straightforwardly extends to cover these cases, at the expense of somewhat more cumbersome notation.

\subsection{Random lattice translates (dithering)} \label{sec:translates}
In \cite{GCD:04,KC:09} a random lattice translate, or dither, known to both transmitter and receiver was included in the lattice code design. The inclusion of a properly chosen random lattice translate builds upon a construction in \cite{EZ:04} and tends to simplify the analysis of MMSE receivers by making the MMSE estimation error independent of the transmitted codeword.

In the setup considered herein we may include such a lattice translate by considering codebooks of the form $\Xset_{r} = (\Lambda_{r} + \bfu) \cap \Rset$ where $\bfu$ is the random lattice translate, possibly dependent on $\rho$ and $r$. This construction allows for the inclusion of the ``mod-$\Lambda$'' nested lattice codes considered in \cite{GCD:04,KC:09}. Note, however, that the specific way in which the mod-$\Lambda$ construction in \cite{GCD:04} maps information messages to codewords, although important from an implementational point of view, is irrelevant to the analysis presented herein as we only consider decoding and not encoding.

The proofs of Theorem~\ref{thrm:main} and Lemma~\ref{lm:dmin} only need to change in that $\bfx, \bfxh \in \Lambda_{r} + \bfu$ replace $\bfx, \bfxh \in \Lambda_{r}$ in order to establish DMT optimality of the regularized lattice decoder given by
$$
\bfxh_{\lat} = \arg \min_{\bfx \in \Lambda_{r} + \bfu}  \|Ê\bfy - \bfH \bfx \|^2 + \|Ê\bfx \|^2_{\bfT} \, .
$$
In particular, the bound in \eqref{eq:metric-transmitted} holds as is, the bound in \eqref{eq:metric-inball} applies to any $\bfxh \in \tfrac{1}{2} \rho^{\frac{\zeta T}{n}}\Bset \cap (\Lambda_{r} + \bfu$), and \eqref{eq:metric-outball} applies to any $\bfxh \notin \tfrac{1}{2} \rho^{\frac{\zeta T}{n}}\Bset$ as before. It follows that regularized lattice decoding is DMT optimal also for designs which include arbitrary chosen random or non-random lattice translates. However, it also follows that no such lattice translate is required for DMT optimality. Still, as argued in \cite{GCD:04,KC:09}, inclusion of a lattice translate could symmetrize the code, and potentially improve the characteristics of the code at finite SNR.

\subsection{Noise generalizations}

It is valuable to point out that Theorem~\ref{thrm:main} is only weakly dependent on the nature of the additive noise. In fact, the only parts of the proof that explicitly depend on the Gaussian assumption, is in the lower bound on the pairwise error probability (PEP) in \eqref{eq:pairwise} and where it is concluded that $\prob{\|Ê\bfw \|^2 \geq \rho^{\delta}} \doteq \rho^{-\infty}$ in Section~\ref{sec:proof}. Thus, for any noise statistics under which $\prob{\|Ê\bfw \|^2 \geq \rho^{\delta}} \doteq \rho^{-\infty}$ and where we may assume a non-zero lower bound on the PEP as in \eqref{eq:pairwise}, the regularized decoder may be shown to at least match the diversity of the (mismatched) ML decoder in \eqref{eq:ml}, i.e., $d_{\lat}(r) \geq d_{\ml}(r)$. In the case of correlated Gaussian noise, the model in \eqref{eq:model} is generally directly applicable after absorbing a noise whitening filter into the channel matrix.

The noise generalization also proves useful when the noise component in \eqref{eq:model} contains self interference, i.e., $\bfw = \bfE \bfx + \bfv$ for some stochastic $\bfE \in \reals^{m \times n}$ and noise $\bfv$. This encompasses the partially coherent scenario when the receiver only knows the channel approximately, in which case $\bfE$ would model the channel estimation error. Under the assumption that $\|\bfE\|\fro^2$ is independent of $\bfx$ and $\rho$, which is typically the case when the channel is estimated using pilots of power proportional to the transmit signal power, and when $\prob{\|Ê\bfE \|^2\fro \geq \rho^{\delta}} \doteq \rho^{-\infty}$ the previous results apply, in spite of the fact that the noise is no longer independent of the transmit signal. In particular, the lower bound of the PEP in \eqref{eq:pairwise} applies straightforwardly by the additive noise alone, and $\prob{\|Ê\bfw \|^2 \geq \rho^{\delta}} \doteq \rho^{-\infty}$ follows by the tail assumption on $\|\bfE\|\fro^2$. We also note that the argument in Section~\ref{sec:proof} does not rely on independence between $\bfx$ and $\bfw$. Thus, the regularized lattice decoder is provably good also in some scenarios involving non-perfect channel state information (CSI) at the receiver.

\subsection{Lower bounds on the diversity} \label{sec:lower-bound}
Finally, consider an arbitrary, continuous, lower bound on the diversity of the ML decoder, i.e., $d_{\ml}(r) \geq \underline{d}_{\ml}(r)$. It is clear that \eqref{eq:outage} holds with $\underline{d}_{\ml}(r)$ in place of $d_{\ml}(r)$. Thus, \eqref{eq:last-limit} and \eqref{eq:last-limit2} also holds with $\underline{d}_{\ml}(r)$ in place of $d_{\ml}(r)$ and it follows that $d_{\lat}(r) \geq \underline{d}_{\ml}(r)$, i.e., that same lower bound applies to the regularized lattice decoder. Naturally, this observation may be of interest in scenarios where the diversity of the ML decoder is discontinuous and/or not explicitly known.

An important special case is where $d_{\ml}(r) = \infty$ over some open interval of $r$. The application of a sequence of continuous  lower bounds may be used to establish that $d_{\lat}(r) = \infty$ over the same interval. Of special interest here is the scenario when lattice decoding of an approximately universal lattice code (e.g., \cite{EKP:06,KC:09}) is restricted to channels not in outage, in which case it follows that $d_{\lat}(r) = d_{\ml}(r) = \infty$. A direct application of this result is given in Section~\ref{sec:mimo-arq}.

\section{Examples} \label{sec:examples}

We proceed by providing a few example scenarios to which the results developed in the previous section are applicable. The examples in Section~\ref{sec:mimo-flat}~,~\ref{sec:mimo-ofdm}~and~\ref{sec:af} are straightforward in the sense that they simply establish a distribution for $\bfH$ in \eqref{eq:model}, to which Theorems~\ref{thrm:main}, \ref{thrm:approx}~and~\ref{thrm:complex} are directly applicable. The example in Section~\ref{sec:mimo-arq} is, however, more involved.

\subsection{The quasi-static MIMO channel} \label{sec:mimo-flat}

The $\nt$-transmit $\nr$-receive antenna quasi-static (flat-fading) MIMO channel commonly given by (c.f.\ \cite{GCD:04})
\begin{equation} \label{eq:multi-antenna-flat}
\bfy_{t}^c = \sqrt{\rho} \, \bfH^c \bfx^c_{t} + \bfw^c_{t} \, , \quad t=1,\ldots,T
\end{equation}
where $\bfH^c \in \complex^{\nr \times \nt}$ has some distribution independent of $\rho$, where $\bfx^c_{t} \in \complex^{\nt}$, $\bfy^c_{t} \in \complex^{\nr}$, and $\bfw^c_{t} \in \complex^{\nr}$, and where $t$ denotes a time index. The channel may be rewritten in the form of \eqref{eq:model} where $\bfx = [\bfx_{1}\tr,\ldots,\bfx_{T}\tr]\tr$ with
$$
\bfx_{t}\tr = [ \Re(\bfx_{t}^c)\tr  \; , \; \Im(\bfx_{t}^c)\tr ]
$$
and where $\Re(\cdot)$ and $\Im(\cdot)$ denotes the real and imaginary part respectively, $\bfw = [\bfw_{1}\tr,\ldots,\bfw_{T}\tr]\tr$ with
$$
\bfw_{t}\tr = [ \Re(\bfw_{t}^c)\tr \; , \; \Im(\bfw_{t}^c)\tr ] \, ,
$$
and
\begin{equation} \label{eq:equiv-channel}
\bfH = \sqrt{\rho} \, \bfI \kron \begin{bmatrix} \Re(\bfH^{c}) & \hspace{-9pt} -\Im(\bfH^c) \\ \Im(\bfH^c) & \hspace{3pt} \Re(\bfH^{c})
\end{bmatrix} \, .
\end{equation}
The channel in \eqref{eq:multi-antenna-flat} is also often written in an equivalent matrix form
\begin{equation} \label{eq:mimo-matrix}
\bfY^c = \sqrt{\rho} \, \bfH^c \bfX^c + \bfW^c
\end{equation}
where $\bfX^c = [\bfx_{1}^c, \ldots, \bfx_{T}^c]$ and $\bfW^c = [\bfw_{1}^c, \ldots, \bfw_{T}^c]$. Under the short-term average input power constraint
\begin{equation} \label{eq:power}
\frac{1}{|\Xset|} \sum_{\bfx \in \Xset} \|Ê\bfX^c \|^2\fro = \frac{1}{|\Xset|} \sum_{\bfx \in \Xset} \|Ê\bfx \|^2 \leq T \, ,
\end{equation}
and an appropriate scaling of $\bfH^c$, the parameter $\rho$ takes on the interpretation of an average signal-to-noise ratio (SNR) per receive antenna  (c.f.\ \cite{ZT:03,GCD:04}).

\subsection{The parallel MIMO channel (MIMO-OFDM)} \label{sec:mimo-ofdm}

A natural extension of the quasi-static MIMO channel is the $\nt \times \nr$ parallel, or MIMO-OFDM, channel. In this setting
\begin{equation} \label{eq:mimo-ofdm}
\bfY^c_{l} = \sqrt{\rho} \; \bfH^c_{l} \bfX^c_{l} + \bfW_{l}^c \, , \quad l=1,\ldots,L
\end{equation}
where $\bfX^c_{l} = [\bfx^c_{l,1},\ldots,\bfx^c_{l,T}]\in \complex^{\nt \times T}$ denotes the complex space-time block codeword transmitted over the $l$th sub-channel in the $T$ time-slots, and where $\bfH^c_{l} \in \complex^{\nr \times \nt}$ is the channel matrix for the $l$th sub-channel. Similar to the flat fading quasi-static channel, it is clear by the linearity of \eqref{eq:mimo-ofdm} that the parallel channel can be rewritten according to \eqref{eq:model}. Coding across the parallel channels is achieved by the appropriate choice of generator matrix $\bfG$. For the rate definition it is conventional to consider one use of \eqref{eq:mimo-ofdm} as $LT$ channel uses.

Naturally, the DMT characteristics of the parallel channel depend of the statistics of $[\bfH^c_{1},\ldots,\bfH^c_{L}]$. In the particular case where $\bfH^c_{l}$ for $l=1,\ldots,L$ represent the OFDM tones for a $Q$-tap i.i.d.\ Rayleigh fading channel, i.e.
$$
\bfH^c_{l} = \sum_{q=0}^{Q-1} \bfHt^c_{q} e^{-i 2 \pi q \frac{l-1}{L}} \, , \quad l=1,\ldots,L,
$$
where $\bfHt^c_{q} \in \complex^{\nr \times \nt}$, $q=0,\ldots,Q-1$, are stochastically independent i.i.d.\ Rayleigh fading taps in the time domain, the maximal diversity gain is $f_{Q}(r)$ where $f_{Q}(r)$ is given by the piecewise linear curve connecting $(k,(Q \overline{n}-k)(\underline{n}-k))$ for $k=1,\ldots,\underline{n}$ where $\overline{n} = \max(\nr,\nt)$ and $\underline{n} = \min(\nr,\nt)$ respectively \cite{MS:05}. Generalizations of this result, to more complicated scenarios, are found in \cite{CB:07}.

Lattice designs for which $d_{\ml}(r) = f_{Q}(r)$ for all $r \in [0,\underline{n}]$ were given in \cite{YBR:06,Lu:08} for particular values of $\nt$ and $L$ and in \cite{EK:07} for the general case of  $n_T, L$. Due to the continuity of $f_{Q}(r)$ we may conclude that low complexity and DMT optimal decoding of these codes is possible, i.e., there exist computationally efficient explicit and DMT optimal transceiver designs for the parallel MIMO channel. The results extend to any statistics under which $d_{\ml}(r)$ is continuous.

\subsection{The amplify-and-forward relay channel} \label{sec:af}

Over the amplify-and-forward (AF) relay channel, one or several relays amplify and retransmit the signal received in previous time-slots, in order to aid the transmission of data from a source to destination. An initial, orthogonal, version of this scenario was in the DMT context studied in \cite{LW:03}. As an example, we here consider another AF protocol, namely the single-antenna single relay non-orthogonal amplify and forward (NAF) protocol proposed in \cite{NBK:04}, operating over a quasi-static channel.  We omit constant transmit power scaling factors for brevity.
One transmission from the source followed by a joint source relay transmission may be modeled according to (c.f.\ \cite{AES:05})
\begin{equation} \label{eq:naf-color}
\bfy^c_{t} = \begin{bmatrix} \sqrt{\rho} h_{1}^c & 0 \\
\rho b h_{2}^c h_{3}^c & \sqrt{\rho} h_{1}^c
\end{bmatrix} \bfx^c_{t} +
 \begin{bmatrix}
0 \\ \sqrt{\rho} b h_{3}^c
\end{bmatrix} w^c_{t} + \bfv^c_{t}
\end{equation}
where $h_{1}^c$, $h_{2}^c$ and $h_{3}^c$ are the complex gains from source to destination, source to relay, and relay to destination respectively. The term $w^c_{t}$ represents the receiver noise at the relay and $\bfv^c_{t}$ the noise at the destination. 
The relay amplification $b$ is in general allowed to depend on $\rho$ and $h_{2}$ and must satisfy
\begin{equation}
|b|^2 \leq \frac{1}{\rho |h_{2}|^2 + 1} \, ,
\end{equation}
in order to meet the relay transmit power constraint.  After noise whitening \eqref{eq:naf-color} becomes equivalent to \eqref{eq:multi-antenna-flat} with
\begin{equation} \label{eq:naf-white}
\bfH^c =
\begin{bmatrix}
h_{1}^c & 0 \\
\frac{\sqrt{\rho}b h_{2}^c h_{3}^c}{\sqrt{\rho |bh_{3}^c|^2+1}} & \frac{h_{1}^c}{\sqrt{\rho |bh_{3}^c|^2+1}}
\end{bmatrix} \, ,
\end{equation}
where one transmission over \eqref{eq:naf-white} corresponds to two channel uses in the definition of the rate. As argued in \cite{YB:07a}, any approximately universal code designed for the $2 \times 2$ quasi-static MIMO channel is able to achieve a diversity gain of $d_{\ml}(r) = (1-r) + (1-2r)^+$, under AWGN noise and i.i.d.\ Rayleigh fading assumptions, provided $b$ is properly selected. This also corresponds to the maximal diversity over the class of linear AF protocols \cite{AES:05}. We see that the AF protocol defines a (somewhat complicated) set of channel statistics, parameterized by $\rho$. It follows directly by the continuity of $d_{\ml}(r)$ that $d_{\lat}(r) = d_{\ml}(r)$ over $r \in [0,1]$.

There are several generalizations of AF protocols to more relays and different relay actions, \cite{AES:05,YB:07b,EVA:09}. General to this setting is that the particular AF protocol determines the statistics of the equivalent channel in \eqref{eq:model}, similar to \eqref{eq:naf-white}.  Lattice designs for some of these generalizations are found in \cite{YB:07a,EVA:09}. The application of Theorem~\ref{thrm:main}, \ref{thrm:approx}  and \ref{thrm:complex} is straightforward to most, if not all, lattice designs in these settings, once the AF protocol is established.  Note that the lattice designs in \cite{YB:07a,EVA:09} provide for approximate universality over these system models, and as a result, $d_{\lat}(r)$ is optimal in these settings.  Note also that even for scenarios where $d_{\ml}(r)$ is not known, it follows from the discussion in Section~\ref{sec:lower-bound} that any continuous lower bound on $d_{\ml}(r)$ applies also to $d_{\lat}(r)$.

\subsection{The $L$-round MIMO-ARQ channel} \label{sec:mimo-arq}

Consider the $L$-round MIMO ARQ setting where, as in \cite{GCD:06}, signaling of the information across the $\nt \times \nr$ quasi-static MIMO channel uses an $L$-round automatic retransmission request (ARQ) protocol that assumes the presence of a noiseless feedback channel conveying one bit of information per use of the feedback channel. During the $l$th round, an $\nt \times T$ code-matrix $\bfX_{l}$ is transmitted where $[\bfX_{1},\ldots,\bfX_{L}] \in \Xset \subset \complex^{\nt \times LT}$, and a decoder $D_{l}$ is applied to decode the fragment $[\bfY_{1},\ldots,\bfY_{l}]$ (c.f.\ \eqref{eq:mimo-matrix} and \eqref{eq:mimo-ofdm}) corresponding to the fragmented code $[\bfX_{1},\ldots,\bfX_{l}] \in \reals^{\nt \times l T}$ with multiplexing gain $r_{l} = r_{1}/l$. The decoder $D_{l}$ either generates an acknowledgment (ACK) in which case a hard decision is made and the transmission of that message terminates, or generates a negative acknowledgment (NACK) in which case another transmission round is requested. The last decoder $D_{L}$ always tries to decode the message. An error is considered only when a message is decoded erroneously. The DMT characteristics of the MIMO-ARQ channel were first considered in \cite{GCD:06} where also the optimal DMT was obtained under two different fading models. We shall for sake of brevity only consider long-term fading where the channel $\bfH^c$ remains constant over the $L$-rounds. We show in what follows how the results obtained herein can be applied to prove DMT optimality of lattice coding and LR-aided linear decoding for the MIMO-ARQ channel, for all $\nr$, $\nt$, $L$ and fading statistics.

To this end, let $\bar{\Aset}_{1}$ denote the event that a NACK is requested in the first round, and let $r_{\max} = \sup \{Êr |Êd_{\out}(r) > 0 \}$ where $d_{\out}(r)$ denotes the optimal DMT for $L=1$, i.e., in the absence of feedback. We assume that $d_{\out}(r)$ is continuous over $r \in [0,r_{\max})$. As in \cite{PKE:09} we consider in parallel a fictitious system where $[\bfX_{1},\ldots,\bfX_{L}]$ is transmitted and where each of the decoders $\Dset_{l}$ operates independently on each of the fragments $[\bfY_{1},\ldots,\bfY_{l}]$, $l=1,\ldots,L$. Let $P_{e,l}(r_{l})$ denote the probability of error of $\Dset_{l}$ in the fictitious system, and let $P_{e}(r)$ denote the overall probability of error at the expected or average multiplexing gain $r$. The work in \cite{PKE:09} provides, based on the work in \cite{GCD:06}, the following sufficient conditions for overall DMT optimality in the MIMO-ARQ setting.
\begin{enumerate}
\item $\prob{\bar{\Aset}_{1}} \doteq \rho^{-\epsilon}$, \; $\epsilon > 0$
\item $P_{e,l}(r_{l}) \dotleq P_{e,L}(r_{L})$, \; $l=1,\ldots,L-1$.
\item $P_{e,L}(r_{L}) \doteq \rho^{-d_{\out}(r_{L})}$.
\end{enumerate}
In brief, optimality follows from the above by observing that
$$
P_{e,L}(r_{L}) \leq P_{e}(r) \leq \sum_{l=1}^L P_{e,l}(r_{l})
$$
which by the second condition implies that $R_{e}(r) \doteq P_{e,L}(r_{L})$. Based on the first condition it may be shown that $r = r_{1}$ (c.f.\ \cite{PKE:09}) and by the third condition it follows that
$$
P_{e}(r) \doteq \rho^{-d_{\out}(r_{L})} =  \rho^{-d_{\out}(\frac{r}{L})}
$$
which corresponds to the maximal ARQ diversity \cite{GCD:06}. The reader is referred to \cite{PKE:09} for a detailed analysis.

Now, let each $\Dset_{l}$ apply regularized lattice decoding, and an ACK-NACK policy similar to \cite{GCD:06,PKE:09} where an ACK is generated if and only if
\begin{equation} \label{eq:ack}
\log \det \! \big( \bfI + \rho \bfH^c (\bfH^c)\hr \big) \geq \frac{x}{l} \log \rho > r_{l} \log \rho \, ,
\end{equation}
for some $x$ such that $r_{1} < x < r_{\max}$. This ACK-NACK policy is independent\footnote{This is a technical requirement for the application of Theorem~\ref{thrm:main} that stems from the fact that we assume the statistics of $\bfH$ in \eqref{eq:model} to be independent of the multiplexing gain of the code applied. Note, however, that the independence is only required in a neighborhood of the target multiplexing gain.} of $r = r_{1}$, provided $r_{1} < x$. Consider now the application of a code where each fragment code is approximately universal. Explicit lattice codes of this type are provided in \cite{EK:09}. As \eqref{eq:ack} implies that the decoders for $l=1,\ldots,L-1$ are only applied to channels not in outage it follows, as explained in Section~\ref{sec:lower-bound}, by the approximate universality of the fragment codes that $P_{e,l}(r_{l}) \doteq \rho^{-\infty}$, for $l=1,\ldots,L-1$. For $l=L$ it follows directly by Theorem~\ref{thrm:main} and \ref{thrm:approx} that $P_{e,L}(r_{L}) \doteq \rho^{-d_{\out}(r_{L})}$. Regarding $\bar{\Aset}_{1}$ it follows by \eqref{eq:ack} that $\prob{\bar{\Aset}_{1}} \doteq \rho^{-d_{\out}(x)}$ where $d_{\out}(x) > 0$ as $x < r_{\max}$, establishing the DMT optimality of the regularized lattice decoder for $r \in [0,r_{\max})$ when applied to the codes proposed in \cite{EK:09}.

We remark that the DMT optimality of lattice coding and decoding for the MIMO-ARQ channel was in fact proven already in \cite{GCD:06}, albeit under the assumption of i.i.d.\ Rayleigh fading and $T \geq \nr + \nt -1$, using a random construction similar to \cite{GCD:04}. The argument presented above extends this result to LR-aided linear decoding, the minimum delay setting ($T = \nt$) and more general fading statistics.

\subsection{Further examples and lattice designs}

The examples given above only constitute a subset of the scenarios to which the main results presented herein are applicable. For instance, ISI channels and generally selective fading channels \cite{CB:07} may be handled similarly to the parallel channel in Section~\ref{sec:mimo-ofdm}. The finite rate feedback scenarios and long term power allocation policies considered in \cite{TS:07} are handled similarly to the MIMO-ARQ channel in Section~\ref{sec:mimo-arq}. Dynamic decode-and-forward (DDF) protocols, where relays decode and forward a received message whenever the relevant channels are not in outage, are also handled similarly to the MIMO-ARQ channel. The results extend to cover orthogonal amplify and forward (OAF) as well as orthogonal and non-orthogonal selection decode and forward (OSDF and NSDF) relay protocols \cite{EVA:09}. Approximately universal distributed codes exist for several such cooperative protocols and scenarios, see e.g., \cite{EVA:09,AES:05,LW:03,NBK:04,JH:06}, and the regularized lattice decoders and their LR-aided linear counterparts achieves the corresponding approximate universality in these settings.

\begin{table}
\renewcommand{\arraystretch}{1.3}
\caption{Lattice dimensionality and references for explicit transceivers in different settings
\label{tab:LatticeDim}}
\begin{center}
\begin{tabular}{||c|c|c||}
\hline \hline
Channel  & $n$      & Lattice source  \\
\hline
$m\times m$ MIMO        &    $2m^2$     &    \cite{EKP:06,ORB:06,KC:09}       \\
\hline
$m\times m$, $L$-tone MIMO-OFDM        &    $2m^2L$     &    \cite{YBR:06,Lu:08,EVA:09}      \\
\hline
$m\times m$, $m$-round MIMO-ARQ         &    $2m^2$     &    \cite{PKE:09}        \\
\hline
$m\times m$, $L$-round MIMO-ARQ (AU)         &    $2m^2 L$     &    \cite{PKE:09}        \\
\hline
$m$-relay OAF   &    $2m$     &    \cite{EVA:09}        \\
\hline
$2$-relay OSDF, NSDF ($r=2$)   &    $32,162 $     &    \cite{EVA:09}         \\
\hline
$m$-relay NAF    &    $8(m-1)$      &    \cite{YB:07a}          \\
                                 &    $8(m-1)^2$      &    \cite{EVA:09}          \\
\hline
$m$-relay DDF, $L$-slots, $m>2$  &    $2m^2L$      &    \cite{EK:09}          \\
\hline \hline
\end{tabular}
\end{center}
\end{table}

Table \ref{tab:LatticeDim} identifies the lattice dimensionality employed by DMT optimal implementations for different channels, as well as refers the reader to explicit descriptions of the designs\footnote{In the case of OAF and $m$-round MIMO-ARQ, DMT optimality is limited to a class of channels.  All relay channels consider single-antenna nodes.}. The potentially very large lattice dimensions faced when decoding such designs makes reduced complexity decoders essential to the successful deployment of these designs.

\section{Conclusion} \label{sec:conclusion}

The work presented an explicit characterization of efficient encoder-decoder structures that meet the fundamental DMT performance limits, and do so for very general channel statistics, dimensions, and models. Specifically, it proved that regularized lattice decoders, and the MMSE-GDFE decoder, provide DMT optimal decoding in its most general form, irrespective of the particular code applied. It also established, for the first time, that computationally efficient LR-aided linear decoders are capable of achieving the entire DMT. The generality of the results obtained lends them applicable to a plethora of pertinent communication scenarios which inherently introduce non-standard channel statistics, code-structure limitations and prohibitively high ML-decoding complexity.

In terms of information theoretic guarantees on error probability performance, the work extended prior state-of-art to a very general setting.
In terms of implementability, the work covered the gap that exists, between the point of proving the existence of non-ML optimal transceivers, and the point of establishing what these transceivers are and how they can be efficiently applied. In terms of complexity guarantees, the work provides worst-case guarantees on the complexity required for DMT optimality.  This is done despite the fact that the employed algorithms are generally known to have unbounded worst-case complexity.

In terms of generality over codes, dimensions and channel statistics, we observe the following: Generality with respect to the codes addresses issues of legacy, and guarantees that the efficient regularized decoder structure will maintain, in most circumstances, the ML decoder DMT performance of the existing code structure.  The generality thus also applies to communication scenarios which place restrictions on the form of the codes applied.  

Generality with respect to channel dimensions is pertinent to computationally demanding scenarios that involve encoding over a large number of degrees of freedom, such as multi-toned OFDM, multi-tap ISI, as well as multi-round MIMO-ARQ and multi-slot DDF channels.  In all the above, error probability performance gains require an increasing number of rounds/slots, which in turn result in linear increases in the problem dimensionality and exponential increases in the ML decoding complexity. The same generality with respect to dimension bypasses issues of channel asymmetry, as well as allows for a unified exposition of the problem.

Finally, generality with respect to fading statistics maintains the pertinent asymptotic guarantees to cases where the underlying fading and noise statistics are not entirely known, specifically to scenarios which inherently introduce hard to characterize channels such as different cooperative relaying protocols, as well as MIMO-OFDM and time-varying channels with arbitrary correlations.

In terms of practicality, the presented transceivers allow for a broad spectrum of rate-reliability-complexity guarantees that result in near-optimal transmission energy, and reduced algorithmic power consumption and delay. Under the requirement for non-exponentially complex decoders, the work also allows for these rate-reliability guarantees in the presence of reduced hardware complexity, such as for example with a minimum number of transmit and receive antennas.  Furthermore the efficient and universal applicability of the transceivers over different system models, allows for further diversification of resources over hybrid channels that near-optimally induce further gains in performance. In terms of future work, the results naturally motivate further joint study into new approximation algorithms and code designs that together yield improved approximation ratios, and better performance in the non-asymptotic regime.

\begin{appendices}

\section{Equivalence of the MMSE-GDFE and the Regularized Lattice Decoder} \label{app:equiv}

By ``completion of squares'' the regularized metric in \eqref{eq:lattice-decoder} may be written according to
\begin{align}
& \| \bfy - \bfH \bfxh \|^2 + \|Ê\bfxh \|_{\bfT}^2 \nonumber \\
= \; & \bfxh\tr \bfH\tr \bfH \bfxh - 2\bfy\tr \bfH \bfxh + \bfy\tr \bfy + \bfxh\tr \bfT \bfxh \nonumber \\
= \; & \bfxh\tr \bfB\tr \bfB \bfxh - 2\bfy\tr \bfF\tr \bfB \bfxh + \bfy\tr \bfF\tr \bfF \bfy + \Gamma \nonumber \\
= \; & \|Ê\bfF \bfy - \bfB \bfxh \|^2 + \Gamma \label{eq:comp-squares}
\end{align}
where $\bfB$ is any matrix for which $\bfB\tr\bfB = (\bfH\tr\bfH + \bfT)$, where $\bfF = \bfB^{-\mathrm{T}} \bfH\tr$ and where
$$
\Gamma = \bfy\tr[\bfI - \bfH(\bfH\tr \bfH + \bfT)^{-1} \bfH\tr]\bfy \geq 0 \, .
$$
As $\Gamma$ does not depend on $\bfxh$ it may be disregarded in the optimization over $\bfxh$, i.e., the regularized lattice decoder may be alternatively expressed as
\begin{equation} \label{eq:mmse-gdfe-app}
\bfxh_{\lat} = \arg \min_{\bfxh \in \Lambda_{r}} \|Ê\bfF \bfy - \bfB \bfxh \|^2 \, .
\end{equation}
Comparing $\bfB$, $\bfF$, and \eqref{eq:mmse-gdfe-app} or \eqref{eq:mmse-gdfe} to the corresponding expressions in \cite{GCD:04}, establishes the equivalence of the regularized decoder and the MMSE-GDFE decoder when $\bfT = \bfI$.

Further, if $\bfxh_{\app}$ is a $C$-approximate solution to \eqref{eq:mmse-gdfe-app}, i.e., if
$$
C \| \bfF \bfy - \bfB \bfxh_{\lat}Ê\|^2 \geq \| \bfF \bfy - \bfB \bfxh_{\app}Ê\|^2 \, ,
$$
for $C \geq 1$, it follows that
$$
C (\| \bfF \bfy - \bfB \bfxh_{\lat}Ê\|^2 + \Gamma )\geq \| \bfF \bfy - \bfB \bfxh_{\app}Ê\|^2 + \Gamma \, ,
$$
which by \eqref{eq:comp-squares} implies that $\bfxh_{\app}$ is also a $C$-approximate solution to \eqref{eq:lattice-decoder}.

\section{Proof of Lemma~\ref{lm:dmin}} \label{app:lemma}

Consider the conditional probability of ML decoder error at multiplexing gain $r$ given that $\bfx \in \Bset$ and $\nu_{r} \leq 1$. As $\nu_{r} \leq 1$ there is $\bfd \in \Bset \cap \Lambda_{r}$, $\bfd \neq \bfZero$, such that $\nu_{r} = \frac{1}{4}\|\bfH\bfd\|^2 \leq 1$ by the definition in \eqref{eq:mindist}. Let $\bfxh = \bfx + \bfd$, and note that $\bfxh \in \Rset \cap \Lambda_{r}$ as $\bfd, \bfx \in \Bset$ and $\bfd,\bfx \in \Lambda_{r}$. In other words, $\bfxh$ is a valid codeword in $\Xset_{r}$. The probability that $\bfxh$ achieves an ML metric which is lower than of $\bfx$ is given by the standard pairwise error probability \cite{TV:05}, i.e.,
\begin{equation} \label{eq:pairwise}
\prob{\bfx \rightarrow \bfxh | \bfx \in \Bset, \bfH } = Q \Big( \tfrac{1}{2} \|Ê\bfH \bfd \|  \Big) \geq Q (1) > 0
\end{equation}
where the last inequality follows by the assumption that $\nu_{r} = \frac{1}{4}\|\bfH\bfd\|^2 \leq 1$, and where $Q(\cdot)$ is the $Q$-function.

Let $\bfxh_{\ml} \in \Xset_{r}$ be the output of the ML decoder. It follows that
\begin{align*}
\prob{\bfxh_{\ml} \neq \bfx} \geq \, & \prob{\bfxh_{\ml} \neq \bfx | \bfx \in \Bset, \nu_{r}Ê\leq 1} \times \\
& \prob{\bfx \in \Bset} \prob{\nu_{r} \leq 1} \, ,
\end{align*}
where we use the independence of $\bfx$ and $\bfH$ (and thus also of $\bfx$ and $\nu_{r}$). By \eqref{eq:pairwise} it follows that $\prob{\bfxh_{\ml} \neq \bfx | \bfx \in \Bset, \nu_{r}Ê\leq 1} \doteq \rho^0$.
By applying the same approximation as in \eqref{eq:ld-size} it may, provided $r > 0$, be shown (c.f.\ \cite{BB:99}) that
$$
\lim_{\rho \rightarrow \infty} \prob{\bfx \in \Bset} = \frac{\vol(\Bset)}{\vol(\Rset)} > 0
$$
when $\bfx$ is uniformly distributed over $\Xset_{r} = \Rset \cap \Lambda_{r}$. This implies that $\prob{\bfx \in \Bset} \doteq \rho^0$. It follows that
$$
\prob{ \nu_{r} \leq 1} \dotleq \prob{\bfxh_{\ml} \neq \bfx}
$$
which is equivalent to \eqref{eq:outage}. \hfill $\square$

\section{Proof of Lemma~\ref{lm:power}} \label{app:power}

Assume that
$$
\prob{ \| \bfH \|^2\fro \geq \rho^x} \geq \rho^{-d_{\ml}(r)}
$$
for sufficiently large $\rho$. It then follows that
\begin{equation} \label{eq:markov}
\expt{ \| \bfH \|^2\fro } \geq \rho^{x-d_{\ml}(r)} \, .
\end{equation}
Thus, if $\expt{\| \bfH \|^2\fro} \dotleq \rho$ it holds that
\begin{equation} \label{eq:app-prob-bound}
\prob{ \| \bfH \|^2\fro \geq \rho^x} \dotleq \rho^{-d_{\ml}(r)}
\end{equation}
for any $x > d_{\ml}(r) + 1$. Let $\bfM = \bfB \bfG$ where $\bfG$ is the code lattice generator and $\bfB\tr\bfB = \bfH\tr\bfH + \bfT$ (c.f.\ Appendix~\ref{app:equiv}). It holds that
$$
\kappa^2(\bfM) = \frac{\lambda_{\max}(\bfM\tr\bfM)}{\lambda_{\min}(\bfM\tr\bfM)}
$$
where $\lambda_{\max}(\bfM\tr\bfM)$ and $\lambda_{\min}(\bfM\tr\bfM)$ denotes the largest and smallest eigenvalues of $\bfM\tr\bfM$. Note that $\bfM\tr\bfM = \bfG\tr \bfB\tr \bfB \bfG$. As $\lambda_{\min}(\bfM\tr\bfM) \geq \lambda_{\min}(\bfG\tr\bfT\bfG) > 0$ and $\lambda_{\max}(\bfM\tr\bfM) \leq \lambda_{\max}(\bfG\tr\bfH\tr\bfH\bfG)+\lambda_{\max}(\bfG\tr\bfT\bfG)$ where $\lambda_{\max}(\bfG\tr\bfH\tr\bfH\bfG) \leq \lambda_{\max}(\bfG\tr\bfG) \|Ê\bfH \|^2\fro$ it follows that
\begin{equation} \label{eq:cond-bound}
\kappa^2(\bfM) \leq \frac{ \lambda_{\max}(\bfG\tr\bfG)\|Ê\bfH \|^2\fro + \lambda_{\max}(\bfG\tr\bfT\bfG)}{\lambda_{\min}(\bfG\tr\bfT\bfG)} \, .
\end{equation}
For $\alpha > \frac{1}{2}x$ it follows by \eqref{eq:cond-bound} that for sufficiently large $\rho$
$$
\|Ê\bfH \|^2\fro \leq \rho^x \quad \Rightarrow \quad \kappa(\bfM) \leq \rho^\alpha \, .
$$
Thus, by \eqref{eq:app-prob-bound} it follows that
$$
\prob{\kappa(\bfM) \geq \rho^\alpha } \dotleq \rho^{-d_{\ml}(r)}
$$
for any $\alpha > \frac{1}{2}(d_{\ml}(r) + 1)$. \hfill $\square$

\end{appendices}

%% Bibliography

\end{document}